\documentclass[superscriptaddress,aps,pra,nofootinbib,twocolumn,notitlepage,10pt]{revtex4-1}
\pdfoutput=1
\usepackage{graphicx}
\usepackage{amsthm}
\usepackage{thmtools}
\usepackage{mathtools}
\usepackage{thm-restate}
\usepackage{amsmath}
\usepackage{amssymb}
\usepackage{comment}
\usepackage{placeins}
\usepackage[caption=false]{subfig}
\usepackage[colorlinks]{hyperref}
\usepackage{tikz}
\usepackage{epstopdf}
\usetikzlibrary{calc,backgrounds}

\usepackage{pgffor}
\usepackage{url}
\usepackage{hypcap}
\usepackage{dsfont}
\usepackage{algorithm}
\usepackage{algorithmic}
\usepackage{soul}
\usepackage{diagbox}

\newcommand{\eq}[1]{Eq.~\hyperref[eq:#1]{(\ref*{eq:#1})}}
\renewcommand{\sec}[1]{\hyperref[sec:#1]{Section~\ref*{sec:#1}}}
\newcommand{\app}[1]{\hyperref[app:#1]{Appendix~\ref*{app:#1}}}
\newcommand{\tab}[1]{\hyperref[tab:#1]{Table~\ref*{tab:#1}}}
\newcommand{\fig}[1]{\hyperref[fig:#1]{Figure~\ref*{fig:#1}}}
\newcommand{\figa}[2]{\hyperref[fig:#1]{Figure~\ref*{fig:#1}#2}}
\newcommand{\figx}[2]{\hyperref[fig:#1]{Figure~\ref*{fig:#1}(#2)}}
\newcommand{\thm}[1]{\hyperref[thm:#1]{Theorem~\ref*{thm:#1}}}
\newcommand{\lem}[1]{\hyperref[lem:#1]{Lemma~\ref*{lem:#1}}}
\newcommand{\cor}[1]{\hyperref[cor:#1]{Corollary~\ref*{cor:#1}}}
\newcommand{\defn}[1]{\hyperref[def:#1]{Definition~\ref*{def:#1}}}
\newcommand{\alg}[1]{\hyperref[alg:#1]{Algorithm~\ref*{alg:#1}}}

\def\ket#1{\mathinner{|{#1}\rangle}}

\newcommand{\be}{\begin{equation}}
\newcommand{\ee}{\end{equation}}
\newcommand{\ba}{\begin{eqnarray}}
\newcommand{\ea}{\end{eqnarray}}

\usepackage{color}
\usepackage{xcolor}

\usepackage{colortbl}
\makeatletter

    \def\CT@@do@color{%
      \global\let\CT@do@color\relax
            \@tempdima\wd\z@
            \advance\@tempdima\@tempdimb
            \advance\@tempdima\@tempdimc
    \advance\@tempdimb\tabcolsep
    \advance\@tempdimc\tabcolsep
    \advance\@tempdima2\tabcolsep
            \kern-\@tempdimb
            \leaders\vrule
                    \hskip\@tempdima\@plus  1fill
            \kern-\@tempdimc
            \hskip-\wd\z@ \@plus -1fill }
    \makeatother

\input{Qcircuit}

\newcommand{\bvec}[1]{\mathbf{#1}}
\renewcommand{\vr}{\bvec{r}}

\begin{document}

\title{Discontinuous Galerkin discretization for quantum simulation of chemistry}

\date{\today}

\author{Jarrod R. McClean}
	\affiliation{Google Research, 340 Main Street, Venice, CA 90291, USA}
\author{Fabian M. Faulstich}
    \affiliation{Hylleraas Centre for Quantum Molecular Sciences, Department of Chemistry, University of Oslo, Oslo, Norway}
\author{Qinyi Zhu}
	\affiliation{Department of Mathematics, University of California, Berkeley, CA 94720, USA}
\author{Bryan O'Gorman}
	\affiliation{Department of Electrical Engineering and Computer Sciences, University of California, Berkeley,  CA 94720, USA}
    \affiliation{Quantum Artificial Intelligence Laboratory, NASA Ames Research Center, Moffett Field, CA 94035, USA}
\author{Yiheng Qiu}
	\affiliation{Department of Physics and Astronomy, University of California, Irvine, CA 92697, USA}
\author{Steven R. White}
	\affiliation{Department of Physics and Astronomy, University of California, Irvine, CA 92697, USA}
\author{Ryan Babbush}
	\affiliation{Google Research, 340 Main Street, Venice, CA 90291, USA}
\author{Lin Lin}
	\affiliation{Department of Mathematics, University of California, Berkeley,  CA 94720, USA}
	\affiliation{Computational Research Division, Lawrence Berkeley National Laboratory, Berkeley, CA 94720, USA}
\date{\today}     

\begin{abstract}
Methods for electronic structure based on Gaussian and molecular orbital discretizations offer a well established, compact representation that forms much of the foundation of correlated quantum chemistry calculations on both classical and quantum computers.  Despite their ability to describe essential physics with relatively few basis functions, these representations can suffer from a quartic growth of the number of integrals.   Recent results have shown that, for some quantum and classical algorithms, moving to representations with diagonal two-body operators can result in dramatically lower asymptotic costs, even if the number of functions required increases significantly. We introduce a way to interpolate between the two regimes in a systematic and controllable manner, such that the number of functions is minimized while maintaining a block diagonal structure of the two-body operator and desirable properties of an original, primitive basis.  Techniques are analyzed for leveraging the structure of this new representation on quantum computers.   Empirical results for hydrogen chains suggest a scaling improvement from $O(N^{4.5})$ in molecular orbital representations to $O(N^{2.6})$ in our representation for quantum evolution in a fault-tolerant setting, and exhibit a constant factor crossover at 15 to 20 atoms. Moreover, we test these methods using modern density matrix renormalization group methods classically, and achieve excellent accuracy with respect to the complete basis set limit with a speedup of 1--2 orders of magnitude with respect to using the primitive or Gaussian basis sets alone.  These results suggest our representation provides significant cost reductions while maintaining accuracy relative to molecular orbital or strictly diagonal approaches for modest-sized systems in both classical and quantum computation for correlated systems.
\end{abstract}
\maketitle

\section{Introduction}
Predicting  properties of both molecular and extended systems from first principles has long been the goal of electronic structure in both correlated classical methods \cite{Helgaker2014}, including new approaches based on tensor networks ~\cite{white1992density,chan2011density,nakatani2013efficient,szalay2015tensor}, and now many approaches based on quantum computing~\cite{Abrams1997,
Ortiz2001,Aspuru-Guzik2005,Wecker2014,Hastings2015,BabbushTrotter,McClean2014,McClean2016,BabbushSparse1,McClean2015}, some of which have even been implemented on experimental devices~\cite{Lanyon2010,Du2010,Peruzzo2013,Shen2015,Santagati2016,OMalley2016,Colless2018,Kandala2017,Dumitrescu2018,Hempel2018}.  A crucial aspect of such simulations is the representation of the problem in a tractable discretization scheme such as a finite-difference method, or (more commonly) a basis set, also known as a Galerkin discretization of the problem.  The selection of the basis influences not only the accuracy of the calculation but also the fundamental scaling of the cost of the simulation as a function of the system size.

The design of basis sets for correlated electronic structure has a long and rich history, packing much of the essential physics and chemistry into very compact representations to exploit the power of existing methods.  While these basis sets have ranged from general purpose to those optimized for individual computations, a common mainstay has been the use of Gaussian-based molecular orbitals~\cite{dunning1989gaussian,kendall1992electron,woon1993gaussian,Helgaker2014}.  Molecular orbitals tend to offer compact representations of correlated problems and also present an energy ordering of orbitals that facilitates further reduction of the space through active space methods.  However, a side effect of this reduction in the number of orbitals is often a Hamiltonian with a quartic number of terms and orbitals that are delocalized in space.

\begin{figure}[t]
    \centering
    \includegraphics[width=0.45\textwidth]{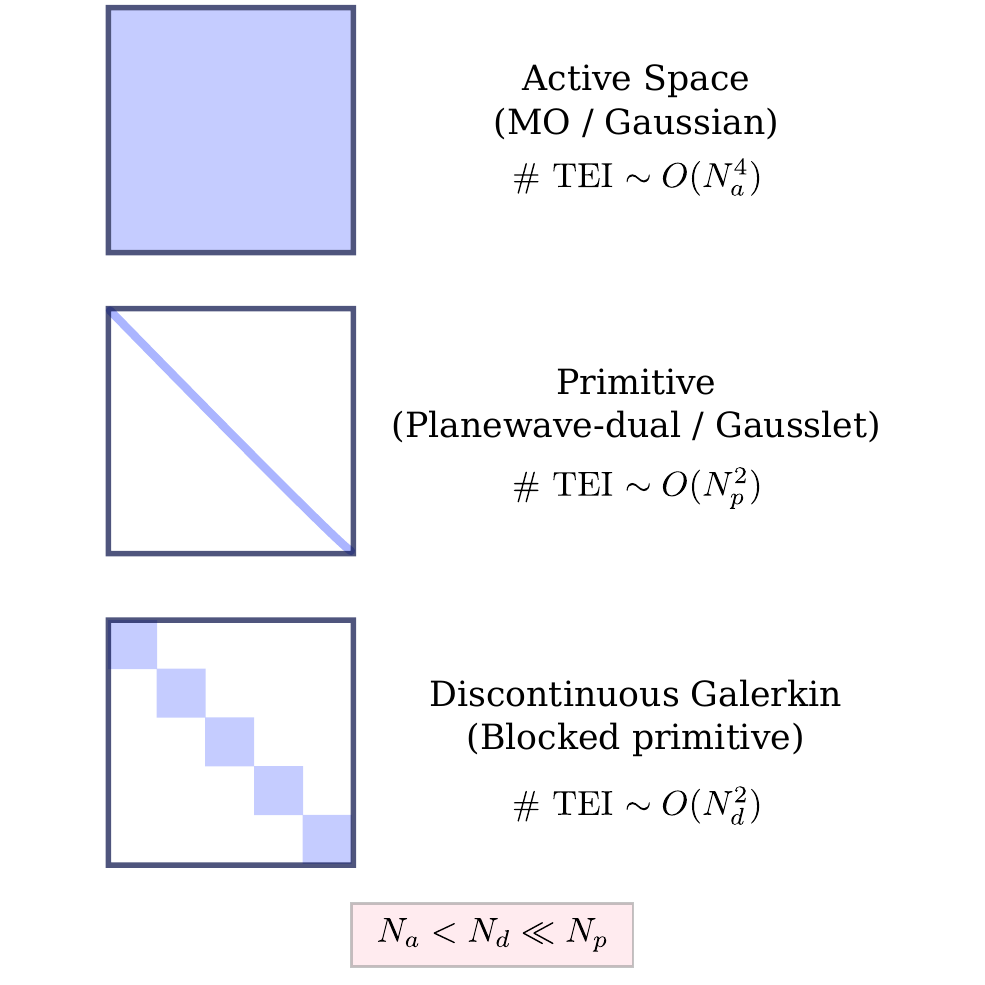}
    \caption{A cartoon schematic of the general objective of this work depicted in terms of the sparsity pattern of the two-electron integrals.  At the top, we depict a dense two-electron integral tensor with $O(N_a^4)$ non-zero two-electron integrals, but relatively few basis functions.  This is often the case with molecular orbital or diffuse Gaussian basis sets.  In the center we depict a diagonal primitive basis set, such as Gausslets, with a relatively high number of basis functions but overall scaling $O(N_p^2)$.  Finally, we depict this for a discontinuous Galerkin (DG) basis set built from a diagonal primitive basis set that interpolates between the two regimes by using fewer basis functions than strictly diagonal sets, while retaining an overall $O(N_d^2)$ scaling through its block diagonal structure.  We note that the form of block diagonal structure is a particular sparsity pattern which depends on the arrangement of indices when considering a matrix form; the text defines this sparsity pattern precisely.
    \label{fig:dg_schematic}}
\end{figure}

The utility of localizing orbitals in both the occupied and virtual space has been recognized in the development of many linear scaling methods~\cite{goedecker1999linear}, such as those based on pair natural orbitals~\cite{ahlrichs1975pno,riplinger2013natural,riplinger2013efficient}.  In these cases the localization helps not only in allowing the screening of some terms in the Hamiltonian, but also reducing the correlations that need to be treated to reach a certain level of accuracy.  However, even the most localized orbitals that can be produced from transformations of a standard basis are often not strictly local in the sense that they have heavy tails that can extend throughout the system to enforce orthogonality between orbitals.  While reasonably compatible with some methods, these tails negatively impact tensor network methods such as the density matrix renormalization group (DMRG), where an area-law entanglement system that is solvable in modest polynomial time becomes a volume law system with exploding bond dimension in a completely delocalized basis~\cite{stoudenmire2017sliced}.  This effect, in combination with consideration of the number of Hamiltonian terms, led to the recent development of Gausslet basis sets~\cite{white2017hybrid,white2019multisliced}, which use properties of wavelet transforms to maintain strict localization, orthogonality, and a diagonal Coulomb operator.  

In the case of quantum computing-based approaches, while basis localization may impact the representational power of an ansatz-based variational approach, the structure and number of terms in the Hamiltonian represent the dominant cost factor \cite{babbush2018low,babbush2018encoding,Berry2019,Motta2018}.  Moreover, some traditional density-based truncations and localizations can be difficult to utilize in quantum computers, due to the need to measure the density and rotate the basis to maximize the benefits.  Recent advances in quantum-computing approaches have shown that Hamiltonians with a diagonal structure with a quadratic number of terms allow algorithms to execute numerically exact real-time dynamics and perform full configuration interaction (assuming a reference state with non-vanishing overlap on the ground state can be prepared) with a cost in terms of basis size that scales as roughly $O(N_p^{1/3})$ in first quantization \cite{BabbushContinuum} or $O(N_p^2)$ in second quantization~\cite{Low2018}.  While the first work was done for periodic systems based on basis sets related to discrete variable representations (DVR) \cite{babbush2018low}, these results apply to all basis sets with these properties.  This is in contrast to the most advanced methods based on Gaussian molecular orbitals on quantum computers, which have costs more like $O(N_a^3)$ in first quantization \cite{BabbushSparse2} or $O(N_a^4)$ in second quantization \cite{Berry2019}.

\begin{figure}[t!]
    \centering
    \begin{tabular}{c c}
        $N_p$  & Primitive basis functions in diagonal basis\\ 
        $N_a$  & Functions in compact active space basis \\ 
        $N_b$  & Blocks in the DG basis \\ 
        $n_\kappa$  & Functions in DG block $\kappa$ \\
        $N_d$  & Total DG functions, $(\sum_\kappa n_\kappa)$
    \end{tabular}
    \caption{Compact description of the notation used throughout the paper in counting basis functions in different representations for the electronic structure problem.  Here discontinuous Galerkin (DG) is the block basis we construct from primitive functions to represent the active space orbitals with a block diagonal Hamiltonian representation.}
    \label{fig:dg_schematic}
\end{figure}

While the scaling advantages of basis sets with diagonal representations appear at reasonable system sizes, reduced representational flexibility can mean the cost of switching representations for equivalent accuracy is still higher than current quantum computers with limited resources can afford. Thus, it is desirable to be able to split the difference between strictly diagonal basis sets and molecular orbitals.
The adaptive local basis set ~\cite{lin2012adaptive,HuLinYang2015a,banerjee2016chebyshev,BanerjeeLinSuryanarayanaEtAl2018,LiLin2019} was recently introduced to achieve this tradeoff in the setting of density functional theory (DFT).
The adaptive local basis is constructed on-the-fly to capture atomic and environmental effects, by solving eigenvalue problems restricted to local domains called the elements. Each basis function is only supported on one element and is discontinuous on the level of the global domain, and the basis functions are ``glued'' together to approximate the continuous electron density using the interior penalty Discontinuous Galerkin (DG) formalism~\cite{Arnold1982,cockburn2000development}. Compared to typical numerical methods for solving DFT, the DG formulation includes extra 
 correction terms to handle the discontinuities of the basis functions at the boundary of the elements. Numerics indicate that roughly 10 to 40 basis functions per atom are often sufficient to achieve chemical accuracy for DFT calculations using pseudopotentials~\cite{hu2015edge,zhang2017adaptive}. 
 
In this work, we introduce a variation of the DG approach for simulation on quantum computers, which has two main differences compared to previous works~\cite{lin2012adaptive,HuLinYang2015a}. First, our target is to model electronic structure at the correlated level instead of the DFT level, and hence we can afford to obtain the best local basis functions, e.g., by starting from molecular orbitals or approximate natural orbitals. This removes a significant step of approximation in the original DG approach due to the solution of local eigenvalue problems with certain artificial boundary conditions (such as the Dirichlet or periodic boundary conditions). Second, our basis functions are represented as the linear combination of primitive basis functions, which means that the basis functions are no longer strictly discontinuous in the continuous space. We demonstrate that the two-body operator can still maintain a block-diagonal structure for efficient quantum simulation, and for simplicity we will still refer to the basis set as the DG basis set. The use of the primitive basis functions removes the need for correction terms that account for the discontinuity, which is similar in spirit to the discrete discontinuous basis projection method~\cite{XuSuryanarayanaPask2018} for DFT calculations.  The DG basis set enables one to interpolate between cheap diagonal representations and compact non-diagonal basis sets with blocks of specified size.

We begin by introducing the standard discretization of the electronic structure problem in a basis, and define precisely what is meant by a diagonal basis representation and strictly localized functions.  We then briefly review basis sets that exhibit spatial locality and diagonal interactions that have already been used in the context of DMRG and quantum computing.  This allows us to highlight the cost advantages of these representations within each approach.  Then, the discontinuous Galerkin approach is introduced as a general framework for maintaining these properties by using those basis sets as a primitive building block.  The block diagonal structure of the resulting Hamiltonians leads us to new cost models based on swap networks for quantum algorithms.  We show that the new approach both maintains accuracy in correlated calculations, and demonstrates a crossover to dramatically lower costs at modest system sizes between 15 and 20 atoms.  We finish with an outlook on how this approach will influence quantum and classical approaches to correlated electronic structure alike.

\section{Discretizing the electronic structure problem}

A crucial aspect of essentially all algorithms for the simulation of electronic systems is discretization of the system into some tractable representation.  This step takes the electronic Hamiltonian that acts in some continuous space, and maps it to a discrete space.  The continuous electronic structure Hamiltonian is given by 
\begin{align}
  \hat{H} = -\sum_i \frac{\nabla_{\mathbf{r}_i}^2}{2} - \sum_{I, j}
\frac{Z_I}{|\mathbf{R}_I - \mathbf{r}_j|} + \sum_{i < j}
\frac{1}{|\mathbf{r}_i - \mathbf{r}_j|} + E_{\text{II}},
\end{align}
where we have assumed atomic units, the Born--Oppenheimer approximation
such that the positions of nuclei $\mathbf{R}_I$ are constants (giving
rise to a constant energy correction $E_{\text{II}}$ for the
nuclear-nuclear interaction), and the $\mathbf{r}_i$ represent the
positions of electrons.  In the case of Galerkin discretizations, where
one chooses a basis set given by some set of orthonormal functions $\{
\chi_i(\mathbf{r}) \}$ (for simplicity we assume a spin-restricted formulation), and enforces
the anti-symmetry of electrons in the operators, one may express the
Hamiltonian in the standard second quantized form. It is possible to
select basis functions from a number of complete sets that allow the
electron-electron interaction to be represented in a way that is
entirely diagonal under a Jordan--Wigner representation in the
computational basis and also exhibits a diagonal property under
matricization.  In this representation, the second quantized Hamiltonian
is given by
\begin{equation}
  \hat{H}^{(p)} = \sum_{\mu,\nu=1}^{N_p} h^{(p)}_{\mu\nu} \hat{b}_{\mu}^{\dagger}
  \hat{b}_{\nu} + \frac12 \sum_{\mu,\nu=1}^{N_{p}} v^{(p)}_{\mu\nu}   
  \hat{n}_{\mu}
  \hat{n}_{\nu},
  \label{eq:Hprimitive}
\end{equation}
where $\hat b_\mu^\dagger$ is a creation operator in the primitive basis, and $\hat n_\mu = \hat b_\mu^\dagger \hat b_\mu$ is a number operator.   We refer to Hamiltonians written in this form as ``diagonal Hamiltonians'', and use such basis sets here as our ``primitive'' basis (associated with the superscript $(p)$ in $\hat{H}^{(p)}$) to efficiently construct compact bases which partially retain this property.

Generally, the coefficients in these expressions are given by the following integrals,
\begin{equation}
\label{eq:ints}
\begin{split}
h^{(p)}_{\mu \nu} &= \int d \vr \ \overline{\chi}_\mu(\vr) \left(
\frac{-\nabla_{\vr}^2}{2} - \sum_{I} \frac{Z_I}{|\mathbf{R}_I - \vr|} \right) \chi_\nu(\vr), \\
v_{\mu \sigma \gamma \nu} &= \int d \vr d \vr' \
\frac{\overline{\chi}_\mu(\vr)\chi_\sigma(\vr)
\overline{\chi}_\gamma(\vr') \chi_\nu(\vr')}{|\vr - \vr'|},
\end{split}
\end{equation}
and the defining property of our primitive basis sets may be written as $v_{\mu \sigma \gamma \nu} \to v_{\mu \nu}^{(p)} \delta_{\mu \sigma} \delta_{\gamma \nu}$, i.e., $v$ becomes a diagonal matrix when we view $(\mu,\gamma)$ as the row index and $(\sigma,\nu)$ as the column index, respectively. Note that the relation between $v_{\mu \sigma \gamma \nu}$ and  $v_{\mu \nu}^{(p)} \delta_{\mu \sigma} \delta_{\gamma \nu}$
is not necessarily an equality: what one requires is that solutions to the Schr\"odinger equation using the two different forms of the interaction
can systematically approach each other as one approaches the complete basis set limit. A grid, defined via finite differences, has the diagonal property, although there is no underlying basis. Here we consider only basis sets, but the
basis functions with the diagonal property are naturally associated with a uniform or non-uniform grid, with typical spacings between grid
points much smaller compared to interatomic distances. 

For a basis set, a sufficient condition for equality of these expressions is that one has functions with strictly disjoint
support, such that (formally) $\chi_\mu(\mathbf{r}) \chi_\sigma(\mathbf{r}) = 0$ for all $\mu \neq \sigma$. However, such a requirement would be unrealistic because upon careful inspection it would imply a simultaneously diagonal kinetic and potential operator if evaluated in the Galerkin formulation in \eq{ints} (in contrast to a finite-difference or overlapping finite-element approach).
Classical finite element methods can produce near-diagonality by allowing overlapping
elements between only spatially neighboring sites, which retains the
favorable scaling, but generically introduces non-orthogonality in the
basis
\cite{PaskSterne2005a,ChenDaiGongEtAl2014,KanungoGavini2017,lehtola2019review}.
Accordingly, methods requiring a return to orthogonality, such as
quantum-computing methods, suffer a transformation to $O(N_p^4)$
terms on orthogonalization, the introduction of unnecessary
orthogonality tails, or both.  More recent developments in classical
discretization have extended this idea of disjoint cells to allow for
truly disjoint basis sets, but at the cost of the introduction of
additional surface terms and discontinuity penalties to reintroduce
physicality into the problem.  These methods are known as Discontinuous
Galerkin (DG) methods~\cite{Arnold1982,cockburn2000development} and were first introduced to electronic structure in the context of density functional theory~\cite{lin2012adaptive,HuLinYang2015a,banerjee2016chebyshev,BanerjeeLinSuryanarayanaEtAl2018}.  They represent a general and rigorous framework for constructing problem representations that have this property of strict (block) locality, and we develop a variation of these methods in this work to achieve this goal for correlated electronic structure methods without the need for the introduction of surface terms.

While the condition of spatial disjointness is a sufficient condition to
obtain a Hamiltonian with $O(N_p^2)$ terms, it is not necessary.  Two
basis sets that have the diagonal property without the equality of the individual matrix elements, and thus without the strict spatial
disjointness property, are the plane wave dual
basis~\cite{babbush2018low} and Gausslet basis~\cite{white2017hybrid}.
The plane wave dual basis is related to periodic sinc functions and
discrete variable representations (DVR) \cite{babbush2018low} and is constructed from an aliased Fourier transform of plane waves for a given box to 
yield a diagonal Hamiltonian. Here the diagonality originates from the plane wave dual functions (also called the periodic sinc functions) that are Lagrange interpolation functions on a uniform grid. It has the advantage that the basis is naturally periodic, and thus is well suited for the treatment of materials and other condensed phase systems.  Moreover, a modification using a truncated Coulomb interaction can enable the treatment of isolated systems.  Two key downsides of this approach are that it inherently reflects a uniform discretization in space of the problem, lacking the ability to adaptively refine sharp features such as the electron-nuclear cusp, and that it has long tails responsible for maintaining the orthogonality of the basis.  This first property is reflected in a large overhead for representing atomic systems to a level of accuracy similar to Gaussians, and the second makes the method difficult to use with geometric entanglement based approaches such as tensor networks.

When integrated with smooth functions, 
Gausslets behave like $\delta$ functions, which results in the diagonal property.   The Gausslets are obtained as linear combinations of arrays of Gaussians using wavelet transforms. Specific moment properties of the wavelet transforms make the Gausslets integrate polynomials up to a certain order like a $\delta$ function. The $\delta$ function property survives under smooth coordinate transformations, so unlike plane wave dual bases, Gausslets can have variable resolutions, with more degrees of freedom near the nuclei to represent the electron-nuclear cusp.   Also in contrast to the plane wave dual basis functions, Gausslets have strong localization characteristics, avoiding long tails.  The lack of tails means these basis sets are naturally suited for tensor network methods and, relatedly, a variational quantum ansatz may have more expressive power at shorter circuit depths.  The ability to more naturally represent inhomogeneous features means the representational overhead is expected to be modest relative to plane wave representations.   

We note, however, that both methods can benefit from the introduction of pseudopotentials, and that the representational power of all single particle basis sets are expected to be limited by the same asymptotic scaling in the limit of a very large number of basis functions.  This means that for very large basis set sizes approaching the complete basis set limit, the representational overheads are expected to be negligible, and the determining factors are the other properties of the basis sets.  Despite this, however, there is considerable interest in treating systems before reaching this limit, where the representational overhead for basis sets can differ considerably.  

In order to meet the demands of compactness one could start from a more generic basis set and use the expressions in Eq.~$\eqref{eq:ints}$ to determine the Hamiltonian.  However, an approach that will prove fruitful here is to use the fact that it is equivalent to start from a complete primitive basis set, such as those above, and project into a compact ``active space'' Hamiltonian.  Gaussian, molecular orbitals, or other active space constructions can been seen as a specific case of the active space we refer to here.

Since the number of primitive basis
functions is typically large compared to the number of electrons,
quantum chemistry calculations, especially at the correlated level, are
often performed using a smaller basis set (e.g. Gaussian basis
functions, atomic orbitals, or molecular orbitals). Let
$\{\varphi_{p}(\vr)\}_{p=1}^{N_{a}}$ be a set of orthonormal single-particle functions;
we refer to them as active space orbitals
(for instance, canonical Hartree--Fock orbitals, or natural orbitals), which can be
expanded using a primitive basis set as
\begin{equation}
  \varphi_{p}(\vr) = \sum_{\mu} \chi_{\mu}(\vr) \Phi_{\mu p}.
  \label{}
\end{equation}
Here $\Phi\in \mathbb{C}^{N_{p}\times N_{a}}$ is a matrix with orthogonal
columns.  While this projection step represents an approximation depending on the nature of the primitive and active bases, it is one that can be systematically controlled and understood by increasing the size of the underlying primitive basis without significantly increasing the size of the resulting block basis we will construct here.  For flexibility and accuracy, we will allow quite general definitions of the active space basis set in this work.  It will pertain to traditional Hartree-Fock canonical orbitals as well as Gaussian basis sets such as the Dunning cc-pVDZ basis set \cite{dunning1989gaussian}, which allows us to use Gaussians with strictly block diagonal properties by going through the primitive basis set using point sampling through the approximate delta function properties of the primitive basis set.  It has been shown previously that for the Gausslet primitive basis set we use, point sampling is extremely accurate due to the delta function property of the basis~\cite{white2017hybrid}. We will also make use of this construction to build hybrid active spaces, where we use a weighted  density matrix from multiple basis sets to define $\Phi$ through its most important natural orbitals.  In particular, we will combine the expressive power of a large primitive basis, such as Gausslets, to capture static correlations through an unrestricted Hartree-Fock defined active space, while including a Gaussian basis empirically refined to express dynamic correlation, e.g. cc-pVDZ through weighting.  In this work we make use of a joined set of density matrices built from a UHF solution $D^{\text{UHF}}$ and Gaussian orbitals $\alpha D^{\text{Gaussian}}$, to form $D = D^{\text{UHF}} + \alpha D^{\text{Gaussian}}$ in the primitive basis, and use a natural orbital truncation of $D$ to define $\Phi$.  Empirically we use a value of $\alpha \approx 0.01$ later in this work when we combine these basis sets, which appears to give an excellent improvement in accuracy.  A more detailed description of this procedure and refinement of the value $\alpha$ is left to a future work.  We term this the hybrid active space approach.  As we only use this hybrid approach in conjunction with the discontinuous Galerkin blocking procedure, we offload concerns about orthogonality in the projected basis to the singular value decomposition (SVD) used in the DG procedure.

Taking as granted the construction of the matrix $\Phi$, we define a rotated set of creation and annihilation operators in the active space as 
\begin{equation}
  \hat{a}^{\dagger}_{p} = \sum_{\mu=1}^{N_p} \hat b^{\dagger}_{\mu} \Phi_{\mu p}, \quad \hat{a}_{p} = \sum_{\mu=1}^{N_p} \hat  b_{\mu}
  \overline{\Phi}_{\mu p},
  \label{eqn:smallbasis_op}
\end{equation}
where $\overline{\Phi}_{\mu p}$ denotes the complex conjugate and we may project the Hamiltonian as
\begin{equation}
  \hat{H}^{(a)} = \sum_{p,q=1}^{N_a} h^{(a)}_{pq} \hat{a}_p^\dagger \hat{a}_q + \frac{1}{2} \sum_{p,q,r,s=1}^{N_a} v^{(a)}_{pqrs} \hat{a}_p^\dagger \hat{a}_q^\dagger \hat{a}_r \hat{a}_s,
  \label{}
\end{equation}
which we refer to as the active space Hamiltonian. 

Generally we see that our primitive basis sets have favorable scaling in number of terms in the Hamiltonian ($O(N_p^2)$ vs $O(N_a^4)$), which often corresponds to better scaling algorithms.  While $N_p$ and $N_a$ have the same asymptotic scaling, for modest sized calculations it is often observed that $N_p \gg N_a$ in order to achieve comparable accuracy.  Here we will seek a way to split the difference between these two regimes by forming a more compact basis that partially retains the diagonal properties, i.e., the resulting Hamiltonian is block-diagonal.

\section{Discontinuous Galerkin discretization}
\label{sec:DG-basis}

At a high level, we construct the block-diagonal basis by fitting
spatially connected blocks of the primitive basis set to the active
basis set, while preserving the properties of the primitive basis set.
We therewith interpolate between the primitive basis set and the active
basis set.  We will refer to the general class of basis sets that
achieve both completeness in some limit and have the diagonal property
as primitive basis sets.

Our goal is to systematically compress the active basis set $\{\varphi_{p}(\vr)\}_{p=1}^{N_{a}}$
into a set of orthonormal basis functions partitioned into elements
(groups), so that basis functions associated with different elements
have mutually disjoint support.  Assume that the index set
$\Omega=\{1,\ldots,N_{p}\}$ can be partitioned into $N_b$
non-overlapping index sets
\begin{equation}
  \mathcal{K} = \{\kappa_1, \kappa_2, \cdots, \kappa_{N_b} \},
\end{equation}
so that $\cup_{\kappa\in \mathcal{K}}\kappa = \Omega$.  
Then the matrix $\Phi$ can be partitioned
into $N_b$ blocks $\Phi_{\kappa} := [\Phi_{\mu p}]_{\mu \in \kappa}$ for
$\kappa\in\mathcal{K}$. Performing the singular value decomposition for $\Phi_{\kappa}$,
\begin{equation}
  \Phi_{\kappa} \approx U_{\kappa} S_{\kappa} V_{\kappa}^{\dagger},
  \label{}
\end{equation}
where $U_{\kappa}$ is a matrix with orthonormal columns corresponding to
the leading $n_{\kappa}$ singular values up to some truncation tolerance
$\tau$, we obtain our compressed basis
\begin{equation}
  \phi_{\kappa,j}(\vr) = \sum_{\mu \in \kappa} \chi_{\mu}(\vr)
  (U_{\kappa})_{\mu,j}.
  \label{}
\end{equation}

\begin{figure}
    \centering
    \includegraphics[width=0.50\textwidth]{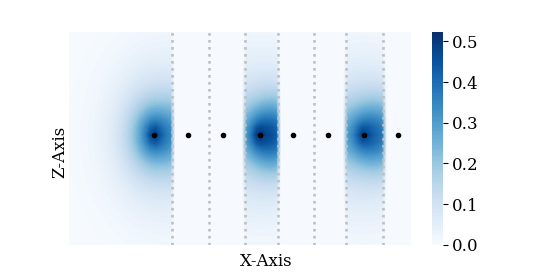}
    \caption{Three DG functions in the X-Z plane, represented in real space. The axes are in units of Bohr and the color intensity represent the amplitude $|\phi_{\kappa,j}|$. Each of the functions is localized to its block, where the block divisions are shown in dotted grey lines. The DG functions are represented by linear combination of plane wave dual basis functions. Each DG function is strictly still a continuous function, but its nodal values (defined according to the center of each plane wave dual function) are only supported within only one block.
    }
    \label{fig:DG-basis}
\end{figure}

The basis set is adaptively compressed with respect to the given set of
basis functions, and are locally supported (in a discrete sense) only on a single index set $\kappa$. In the absence of SVD truncation, we clearly
have $\text{span} \{\varphi_{p}\}\subseteq \text{span}
\{\phi_{\kappa,j}\}$.  We refer to this basis set
$\{\phi_{\kappa,j}\}$ as the DG basis set. Note that each DG basis function $\phi_{\kappa,j}$ is a linear combination of primitive basis functions which are themselves continuous, so $\phi_{\kappa,j}$ is also technically continuous in the real space. 
In fact, $\phi_{\kappa,j}$ 
might not be locally supported in the real space if each primitive basis function
$\chi_{p}$ is delocalized. When the primitive basis functions are localized, $\phi_{\kappa,j}$ can be very close to a discontinuous function. (See Fig.~\ref{fig:DG-basis} for an example.) When computing the projected Hamiltonian, we do not need to evaluate the surface terms in the DG formalism.  If we form a block diagonal matrix
\begin{equation}
  U = \text{diag}[U_{1},\ldots,U_{N_b}],
  \label{}
\end{equation}
the total number of basis functions is thus
$N_{d}:=\sum_{\kappa\in\mathcal{K}} n_{\kappa}$. We remark that the number
of basis functions $n_{\kappa}$ can be different across
different elements. 

To facilitate the complexity count below we may,
without loss of generality, assume that $n_{\kappa}$ is a constant and that
$N_{d}=N_{b} n_{\kappa}$.  
Then we have defined a new set of creation and annihilation operators
\begin{equation}
  \hat{c}^{\dagger}_{\kappa,j} = \sum_{\mu} \hat{b}^{\dagger}_{\mu}
  (U_{\kappa})_{\mu j}, \quad \hat{c}_{\kappa,j} = \sum_{\mu} \hat{b}_{\mu}
  (\overline{U}_{\kappa})_{\mu j},
  \label{eqn:dalb_op}
\end{equation}
with $\kappa=1,\ldots,N_{b}$ and
$j=1,\ldots,n_{\kappa}$ that correspond to the DG basis set.

Unlike Eq.~\eqref{eqn:smallbasis_op}, the basis set rotation in
Eq.~\eqref{eqn:dalb_op} is restricted to each element $\kappa$. 
We readily obtain the projected Hamiltonian in the DG basis as
\begin{align}
  \hat{H}^{(d)} & = \sum_{\kappa,\kappa';j,j'} h^{(d)}_{\kappa,\kappa';j,j'}
  \hat{c}^{\dagger}_{\kappa,j} \hat{c}_{\kappa',j'}\nonumber\\
  & +
  \frac12 \sum_{\kappa,\kappa';i,i',j,j'}
  v^{(d)}_{\kappa,\kappa';i,i',j,j'}
  \hat{c}^{\dagger}_{\kappa,i} \hat{c}^{\dagger}_{\kappa',i'}
  \hat{c}_{\kappa',j'} \hat{c}_{\kappa,j}.
  \label{eqn:H_DG}
\end{align}
The matrix elements are
\begin{equation}
  h_{\kappa,\kappa' ; j,j'}^{(d)} = \sum_{\mu \nu} (\overline{U}_{\kappa})_{\mu j}
  h^{(p)}_{\mu \nu} (U_{\kappa'})_{\nu j'},
  \label{}
\end{equation}
and
\begin{equation}
  v_{\kappa,\kappa' ; i,i',j,j'}^{(d)} = 
  \sum_{\mu \nu } (\overline{U}_{\kappa})_{\mu i}
  (\overline{U}_{\kappa'})_{\nu i'} v^{(p)}_{\mu \nu}
  (U_{\kappa})_{\mu j} (U_{\kappa'})_{\nu j'}.
  \label{}
\end{equation}
In general, the one-body matrix $h^{(d)}$ can be a full dense matrix, but
the two-body tensor $v^{(d)}$ always takes a ``block diagonal'' form in
the following sense (it has a specific sparsity pattern). In principle, the two-body interaction in the DG
basis set should take the form
\begin{equation}
\frac12 \sum_{\kappa,\kappa',\lambda,\lambda';i,i',j,j'}
  v_{\kappa,i;\kappa',i';\lambda,j;\lambda',j'}
  \hat{c}^{\dagger}_{\kappa,i} \hat{c}^{\dagger}_{\kappa',i'}
  \hat{c}_{\lambda',j'} \hat{c}_{\lambda,j}.
\end{equation}
Compared to Eq.~\eqref{eqn:H_DG}, we find that
\begin{equation}
v_{\kappa,i;\kappa',i';\lambda,j;\lambda',j'} =
v^{(d)}_{\kappa,\kappa';i,i',j,j'} \delta_{\kappa\lambda}
\delta_{\kappa'\lambda'}.
\end{equation}
In other words, $v$ can be viewed as a block diagonal matrix with
respect to the grouped indices $(\kappa\kappa',\lambda\lambda')$.
\begin{figure}
    \centering
    \includegraphics[width=0.4\textwidth]{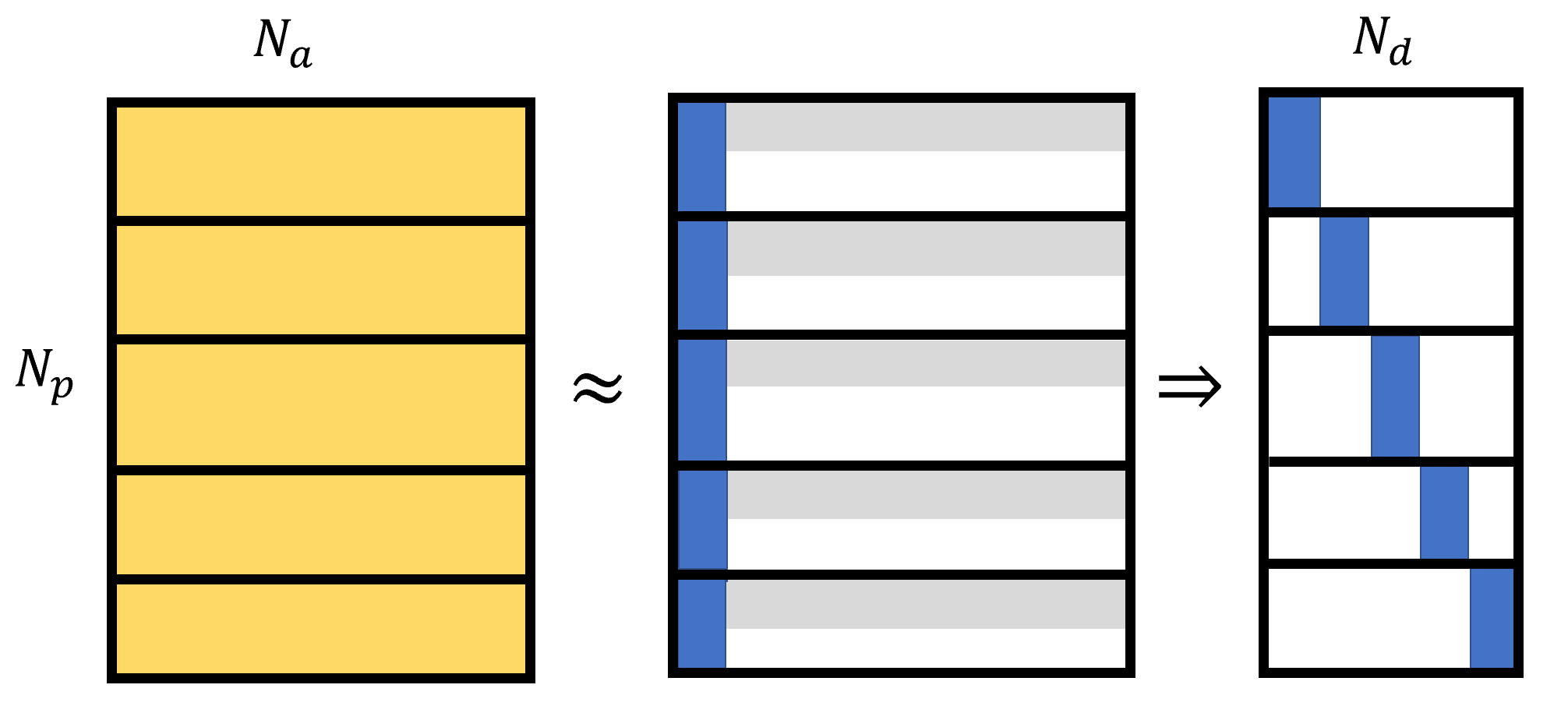}
    \caption{A schematic illustration of the compression process of delocalized active space basis functions into DG basis functions.  Beginning with a matrix representing the projection of a primitive basis onto a chosen active basis (Left), with the primitive basis grouped into blocks represented by rows here.  Those blocks are then reduced by a singular value decomposition (Center), which finally leads to the DG basis that has a block diagonal two-electron integral representation (Right).}
    \label{fig:DGfactorize}
\end{figure}

We remark that the convergence of the DG basis set is independent of the
choice of the primitive basis set so long as the primitive basis has
sufficient degrees of freedom to form a good approximation to the active space functions of interest.  At the end of this adaptive procedure,
we expect the number of elements in the Hamiltonian to scale as $O(N_b^2
n_{\kappa}^4)$.  However, as we expect the number of basis functions required to reach a fixed accuracy within a block (i.e., $n_{\kappa}$) to be bounded by a constant as system size grows,  and the scaling ,with system size becomes $O(N_d^2)$.  We substantiate the rapid asymptotic convergence of $n_\kappa$ for real systems later in this work, however, simple arguments from spatial locality and basis set completeness lead to the same conclusion.

\section{Quantum simulation with a DG basis}
Here we introduce the methods used to exploit the properties of the DG basis on a quantum computer for evolution under the Hamiltonian.  We first describe a method that implements evolution using a Suzuki-Trotter decomposition through swap networks, taking advantage of recent advances in quantum swap networks.  An alternative Trotter approach based on low-rank decompositions is detailed in Appendix~\ref{app:low_rank}.  We then describe a method for time evolution based on the linear combinations of unitaries approach, which will provide the required background for determining the degree of advantage for using DG basis sets for fault-tolerant quantum computations of chemistry.

\subsection{Swap networks for block diagonal Hamiltonians}\label{sec:swap}
In the first work using a strictly diagonal basis in quantum computing for chemistry \cite{babbush2018low}, the ability for quantum computers to perform fast Fourier transforms on quantum wavefunctions was exploited to capitalize on the representational advantages of being in either the plane wave basis or its Fourier-transformed dual.  That method was originally restricted to Hamiltonians with that particular structure in the coefficients, similar to split-operator Fourier transform methods used in classical simulation of quantum systems. However, it was soon realized that the structure of any diagonal Hamiltonian could be similarly exploited.  This generalization used a linear, fermionic swap network to achieve perfect parallelization of a Trotter step with depth that scales linearly in the number of orbitals \cite{kivlichan2018quantum}, even when gates are restricted to act on nearest neighbors of a line of qubits.

\begin{figure}[t]
    \centering
    \includegraphics[width=0.45\textwidth]{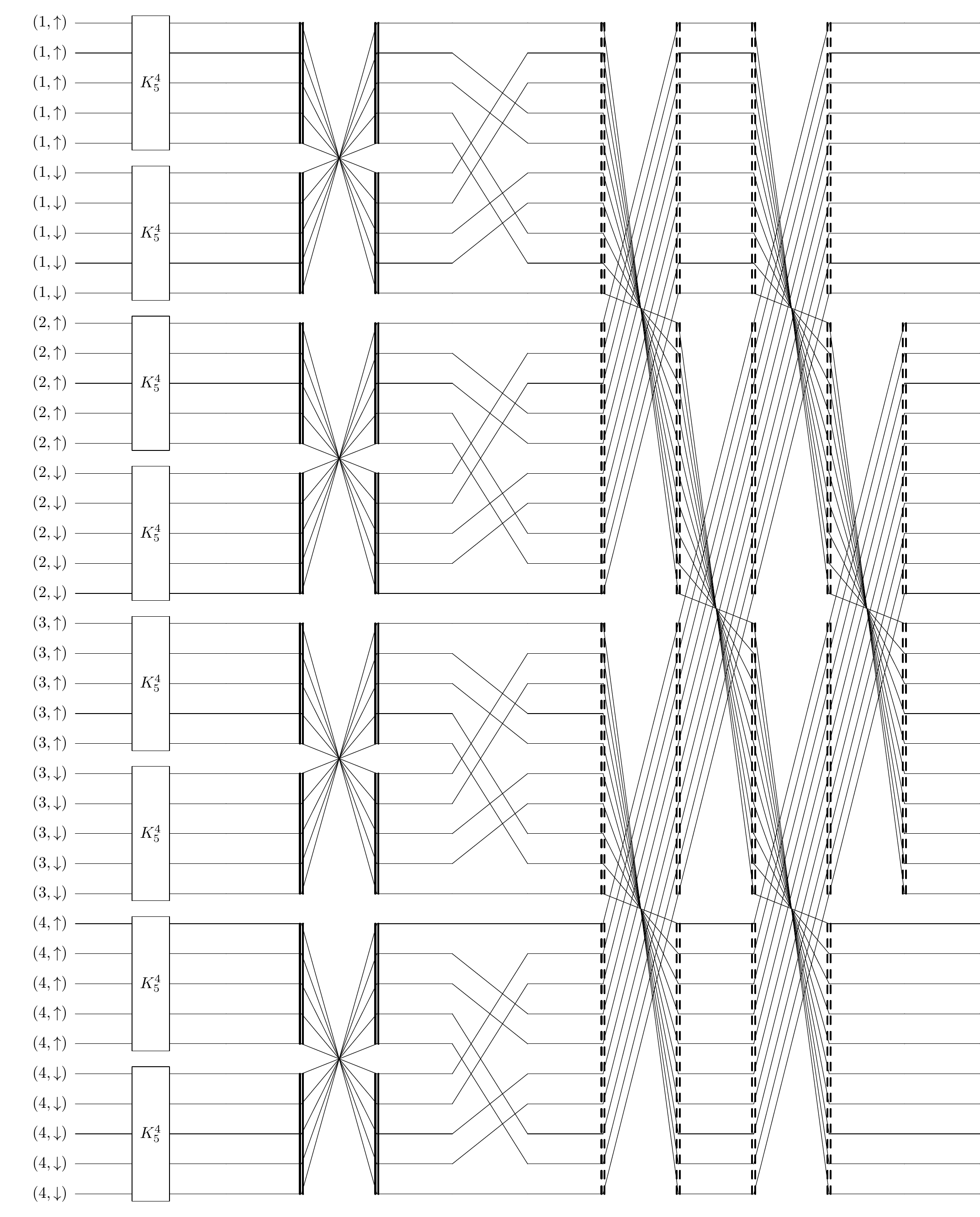}
    \caption{
Acquaintance strategy for block-diagonal Hamiltonian with $N_b=4$ and $n_{\kappa}=10$.
``$K^4_n$'' indicates a 4-complete swap network on $n$ qubits, i.e., one
that acquaints the $\binom{n}{4}$ subsets of $4$ qubits with each other.
The other gates are double bipartite swap networks, explained in
Figures~\ref{fig:double-bipartite}
and~\ref{fig:double-bipartite-balanced}.
\label{fig:block-diagonal-acquaintance-strategy}}
\end{figure}

Fermionic swap networks are analogous to sorting networks from traditional computer science except built upon the primitive of the fermionic swap operation,
\begin{align}
& \hat f^{pq}_{\text{swap}} = 1 + \hat a_p^\dagger \hat a_q + \hat a_q^\dagger \hat a_p - \hat a_p^\dagger \hat a_p - \hat a_q^\dagger a_q, \\
& \hat f^{pq}_{\text{swap}} \hat a_p^\dagger (\hat f^{pq}_{\text{swap}})^\dagger = \hat a_q^\dagger,
\end{align}
where $\hat f^{pq}_{\text{swap}}$
is the fermionic swap that swaps the labeling of modes. The fermionic
swap operation was introduced in \cite{Bravyi2002} and also studied in the
context of tensor networks. The difference between such a swap and a
traditional swap is by swapping fermionic modes instead of assignments
to qubits, non-local parity strings used to enforce the fermionic anti-commutation relations can be avoided. 

The basic idea of the linear swap network is to fermionic swap all neighboring qubits, interact them with their current neighbors, and repeat until all qubits have interacted with each other.  Since the introduction and use of these linear fermionic swap networks in quantum algorithms, they have been generalized for use in non-diagonal Hamiltonians with some overhead.  For example, the quantum chemistry Hamiltonian can be decomposed into a sum of diagonal Hamiltonians (each in a rotated basis) using techniques similar to Cholesky or density fitting methods, where each diagonal Hamiltonian can then be implemented in sequence \cite{Motta2018}.  We show how to use this method in the DG representation in Appendix~\ref{app:low_rank}.

For the general Hamiltonian in quantum chemistry, which is non-diagonal, a generalized swap network that works directly with such Hamiltonians was developed~\cite{ogorman2019generalized}. This network implements time steps for generic $O(N_a^4)$ Hamiltonians in a time that scales as $O(N_a^3)$, and we take advantage of it here with specializations for the block diagonal structure. 
To implement a Trotter step of the Hamiltonian, the swap network dynamically updates the mapping from qubits to orbitals so that for each term in the Hamiltonian, the involved orbitals are mapped to adjacent qubits.
We say that a swap network ``acquaints'' a set of orbitals when it brings them together at some point in this way, and represent that point by an empty box in the circuit diagrams, which acts as a placeholder for the logical gate to be executed there.
Prior work utilizing swap networks has applied them to two extremal regimes with respect to the structure of the two-electron terms in the Hamiltonian: the strictly diagonal case, which can be implemented with $O(N_p)$ depth~\cite{kivlichan2018quantum}; and the fully general case, which can be implemented in $O(N_a^3)$~\cite{ogorman2019generalized}.
Here we show how to interpolate between these to achieve $O(N_b n_{\kappa}^3) = O(N_d n_{\kappa}^2)$ depth for block-diagonal Hamiltonians.
(For simplicity, in this section we will assume that all blocks have the same size $n_{\kappa}$, but the techniques generalize in a straightforward way to non-uniform block sizes.)

We focus on how to implement the quartic terms in the Hamiltonian (i.e., two-electron terms involving four distinct spin orbitals).
The lower-order terms can be addressed with negligible additional resources by incorporating them into the quartic terms.
The quartic terms in the block-diagonal Hamiltonian satisfy the following properties:
\begin{enumerate}
    \item Two orbitals are from one block $\kappa$ and two orbitals are from another block $\kappa'$ (or all four from the same block when $\kappa = \kappa'$). 
    \item The orbital spins have even parity (i.e., all up, all down, or two and two).
\end{enumerate}
We will exploit both of these properties in constructing our swap network, which uses primitives originally designed for implementing unitary coupled cluster~\cite{ogorman2019generalized}.

Figure~\ref{fig:block-diagonal-acquaintance-strategy} shows the overall swap network.
Initially, the orbitals are arranged on the line in lexicographical ordering; only the block index $\kappa$ and spin are indicated for concision.
The logic of the strategy is as follows:
\begin{enumerate}
\item
The first layer acquaints all sets of four spin orbitals within each block in which all four orbitals have the same spin.
This is achieved by a ``4-complete'' swap network on each half-block of orbitals, denoted by $K^{4}_{n_{\kappa}/2}$ because the sets of orbitals it acquaints correspond to the edges of a complete 4-uniform hypergraph; it has depth $O(n_{\kappa}^3)$. Note that the edges of the complete $k$-uniform hypergraph $K^k_n$ on $n$ are the $\binom{n}{k}$ sets of $k$ vertices. The ``uniform'' qualifier indicates that all of the hyperedges have the same number of vertices.
See~\cite{ogorman2019generalized} for details.
\item
The second layer acquaints all sets of four spin orbitals within each block in which two orbitals have spin up and the other two have spin down.
This is achieved by a ``double bipartite'' swap network on each block in depth $O(n_{\kappa}^3)$; see Figure~\ref{fig:double-bipartite}. 
\item
The third layer permutes, in $O(n_{\kappa})$ depth, the orbitals within each block in preparation for the inter-block acquaintances to follow.
\item
The rest of the strategy consists of $N_b$ alternating layers that acquaint pairs of parts.
In each layer, each block of qubits is paired up with an adjacent one and a ``balanced double bipartite'' swap network is executed on the pair of blocks; see Figure~\ref{fig:double-bipartite-balanced}.
Each balanced double bipartite swap network acquaints the sets of four orbitals containing two from each block and with even (``balanced'') spin parity. 
This also has the effect of swapping the blocks, so overall a balanced double bipartite swap network is applied to every pair of blocks.
Each double bipartite swap network has depth $O(n_{\kappa}^3)$.
\end{enumerate}
Overall, the depth is $O(N_b n_{\kappa}^3)$, dominated by the latter swap networks that effect the inter-block interactions.
The components of this approach are explained in more detail in Appendix~\ref{sec:swap-details} and an alternative approach that may have asymptotic advantages in some regimes is discussed in Appendix~\ref{app:low_rank}.

\subsection{LCU approaches for simulation}
\label{sec:LCU_cost}
The quantum simulation algorithms discussed in the previous section and in Appendix~\ref{app:low_rank} are useful for implementing Trotter steps of the chemistry Hamiltonian. Such Trotter steps can be repeated to perform time-evolution for modeling dynamics or for preparing eigenstates via the phase estimation algorithm \cite{Kitaev1995,Abrams1999}, but they can also serve as an ansatz for composing quantum variational algorithms \cite{McClean2015,Wecker2015a}. This in conjunction with these Trotter steps requiring only minimal (linear) connectivity makes them attractive algorithms for near-term quantum computing. However, within cost models appropriate for a fault-tolerant quantum computer, Trotter steps are not the most competitive technique for chemistry simulation. In such a cost model, the key resource to minimize is the number of T gates required by the algorithm. This is because in most practical error-correcting codes (e.g., the surface code), T gates require many physical qubits for their distillation and are orders of magnitude slower to implement than Clifford gates.

Currently, the lowest T complexity quantum algorithms for simulating chemistry are all based on linear combinations of unitaries (LCU) methods \cite{Childs2012}. This family includes Taylor series methods \cite{Berry2015} (applied to chemistry in \cite{BabbushSparse1,BabbushSparse2}), qubitization \cite{Low2016} (applied to chemistry in \cite{Low2016,BabbushSpectra,Berry2019}) and Hamiltonian simulation in the interaction picture \cite{Low2018} (applied to chemistry in \cite{Low2018,BabbushContinuum}). What all LCU methods have in common is that they involve simulating the Hamiltonian from a representation where it can be accessed as a linear combination of unitaries,
\begin{equation}
\label{eq:lcu}
H = \sum_{\ell=1}^{L} \omega_\ell \, U_\ell, \qquad \lambda = \sum_{\ell=1}^L \left| \omega_\ell \right |,
\end{equation}
where $U_\ell$ are unitary operators, $\omega_\ell$ are scalars, and $\lambda$ is a parameter that determines the complexity of these methods. The second quantized Hamiltonians discussed in this paper satisfy this requirement once mapped to qubits (e.g.,~under the Jordan--Wigner transformation) since strings of Pauli operators are unitary. 

LCU methods perform quantum simulation in terms of queries to two oracle circuits defined as
\begin{align}
\textsc{select} \ket{\ell} \ket{\psi} & \mapsto \ket{\ell} U_\ell \ket{\psi},\\
\textsc{prepare} \ket{0}^{\otimes \log L} & \mapsto \sum_{\ell=1}^L \sqrt{\frac{\omega_\ell}{\lambda}} \ket{\ell},
\end{align}
where $\ket{\psi}$ is the system register and $\ket{\ell}$ is an ancilla register which usually indexes the terms in \eq{lcu} in binary and thus contains $\log L$ ancillae. Up to log factors in precision and other system parameters, LCU methods can perform time-evolution with gate complexity scaling as
\begin{equation}
{O}\left( \left(C_S + C_P\right)\lambda\, t\right),
\end{equation}
where $C_S$ and $C_P$ are the gate complexities of  $\textsc{select}$ and  $\textsc{prepare}$ respectively, and $t$ is time.

To implement the LCU oracles one must be able to coherently translate the index $\ell$ into the associated $U_\ell$ and $\omega_\ell$. In the quantum chemistry context the $U_\ell$ are related to the second quantized fermion operators (e.g., $\hat a^\dagger_p \hat a^\dagger_q \hat a_r \hat a_s$) and the $\omega_\ell$ are related to the molecular integrals. While the $U_\ell$ have a structure that is straightforward to unpack in a quantum circuit (see \cite{BabbushSpectra,Berry2019} for explicit implementations), the $\omega_\ell$ are typically challenging to compute directly from this index (unless one pursues the highly impractical strategy of computing the integrals on-the-fly, as in \cite{BabbushSparse1}). As a consequence, with state-of-the-art LCU methods for simulating chemistry the primary bottleneck has been implementation of $\textsc{prepare}$ rather than $\textsc{select}$ \cite{BabbushSpectra,Berry2019}. 

In the most performant LCU approaches for chemistry (see \cite{BabbushSpectra} for plane waves and \cite{Berry2019} for arbitrary basis sets), $\textsc{prepare}$ is implementing (and bottlenecked) by using a data-lookup routine referred to as QROM (quantum read-only memory) to load the $\omega_\ell$ values into superposition. Using this approach, the cost of $\textsc{prepare}$ is a function of the number of unique coefficients $\omega_\ell$ in the Hamiltonian. Using the QROM of  \cite{Lowpreparation} (which improves on the original concept from \cite{BabbushSpectra}), one can (up to log factors) implement $\textsc{prepare}$ with T complexity scaling as ${O}(L/g + g)$ where $L$ is the number of molecular integrals and $g$ is a free parameter so long as at least $g$ dirty ancilla are available during the implementation of $\textsc{prepare}$. As $\textsc{prepare}$ acts only on the ancilla register, there are typically at least $N_d$ dirty ancilla available (from the $\ket{\psi}$ register). Thus (assuming $L > N_d^2$) the scaling is ${O}(L / N_d)$ without ancilla or ${\cal O}(\sqrt{L})$ with ${ O}(\sqrt{L})$ ancilla (often a reasonable tradeoff within error-correction).

When simulating in a molecular orbital basis with $L = {O}(N_a^4)$, one can (up to log factors) evolve the Hamiltonian for some time $t$ and achieve T complexity ${O}(N_a^2 \lambda t)$ \cite{Berry2019}. If a straightforward extension of the approach in \cite{Berry2019} is applied to the DG basis Hamiltonian with $L = n_\kappa^4 N_b^2 $ then this scaling would be reduced to ${O}( n_\kappa^2 N_b \lambda t)$. However, if one can index the $\omega_\ell$ in a way that exploits symmetries in the coefficients, then one can further reduce the effective value of ``$L$'' in both of these expressions. For instance, in \cite{BabbushSpectra} it is recognized that while there are $N_p^2$ coefficients $\omega_\ell$ in the Hamiltonian of \eq{Hprimitive} (one for each value of $\nu$ and $\mu$), there are only $N_p$ \emph{unique} coefficients due to the translational invariance of the Coulomb operator (essentially, one for each unique value of the index $\mu - \nu$), and the cost of the algorithm is reduced accordingly. As a rough rule of thumb, the scaling becomes ${O}(J / g + g)$ where $J$ can be thought of as the number of unique scalars needed to completely describe the Hamiltonian without recomputing the molecular integrals. In \cite{Berry2019}, it is shown that because only $J = {O}(N_a^3)$ numbers are needed to describe the low rank factorized Coulomb operator in an arbitrary basis \cite{Motta2018}, one can improve the scaling to ${O}(N_a^{3/2} \lambda t)$. By tailoring such techniques to symmetries such as periodic crystalline symmetry or other redundancies in the DG Hamiltonian (e.g., those exploited in the low rank representation of Appendix~\ref{app:low_rank}) one can also further improve the ${\cal O}(n_\kappa^2 N_b \lambda t)$ scaling.

Finally, we note that the $\lambda$ value associated with the molecular orbital basis Hamiltonian is likely larger than the $\lambda$ value associated with the DG Hamiltonian, and we quantify the extent to which that is the later in this paper. This is yet another way in which the DG representation should lead to even more efficient implementation of LCU methods for chemistry simulation. Note also, that the value of $\lambda$ is important for the scaling of quantum simulation in a near-term cost model as well. In particular, for quantum variational algorithms the number of measurements (corresponding to the number of circuit repetitions) required to estimate the energy of a Hamiltonian to within error $\epsilon$ scales as $O(\lambda^2 / \epsilon^2)$ \cite{Rubin2018}.

\section{Numerical results}
To understand the performance of different discretization schemes, we examine the costs and associated accuracy for each of the methods, using hydrogen chains of increasing lengths as test systems.  We show that the crossover point occurs around 15 to 20 atoms, above which the DG representation with the block-diagonal Hamiltonian structure has significant lower costs.  

The subsequent calculations are twofold. We begin with numerical
simulations designed to exhibit the expected crossover behavior.  We
investigate hydrogen chains of increasing length (up to H$_{30}$) in a
Gaussian cc-pVDZ basis~\cite{dunning1989gaussian} (the active space to
be fit), where we choose the primitive
diagonal basis set to be plane wave dual functions with refinement built
to match the accuracy of the Gaussian basis set to a specified tolerance
at the level of density functional theory with the
Perdew--Burke--Ernzerhof (PBE) exchange-correlation
functionals~\cite{perdew1996generalized}. 
We are not aware of any electronic structure software package using a plane wave
basis set for practical all-electron DFT calculations. Hence we use the
pseudopotential formulation 
based on the ONCV
pseudopotential~\cite{hamann2013optimized}. 
As we fit to the span of the
basis set itself, rather than a density matrix generated with a fixed
level of theory, these results should be representative of performance
across different levels of theory including, but not limited to,
correlated methods.  This is done to observe the crossover without
extrapolation, as by the size the crossover occurs, the more accurate
correlated calculations become prohibitively expensive using traditional
classical methods.

In the second set of calculations, we instead fit to a natural orbital active space using a Gausslet basis set as the primitive diagonal basis set \cite{white2017hybrid,white2019multisliced} to demonstrate that this representation maintains accuracy in correlated calculations.  This is possible as the properties of a Gausslet basis allow accurate DMRG calculations for comparison on these systems.  Similar DMRG calculations are expected to more expensive in basis sets with heavy tails, such as the plane wave dual basis, which makes the one-particle Hamiltonian a dense matrix, and result in excessive bond dimension requirements for accurate descriptions.  These calculations exhibit the versatility of this method and its ability to achieve accuracy, even in a correlated active space built fit to a highly non-local active space. Furthermore, we demonstrate that by fitting the DG basis simultaneously to two active basis sets (molecular orbitals at the UHF level and the cc-pVDZ basis functions), the DG basis set can 
obtain accurate results when compared to quantum Monte Carlo calculations at the complete basis set limit, while achieving a 1-2 orders of magnitude reduction of computational cost over the primitive Gausslet basis or the Gaussian basis set.

\subsection{Scaling crossover in a DG plane wave dual basis}
\label{sec:DG-DFT-Calc}
\begin{figure}[t!]
    \centering
    \includegraphics[width=0.40\textwidth]{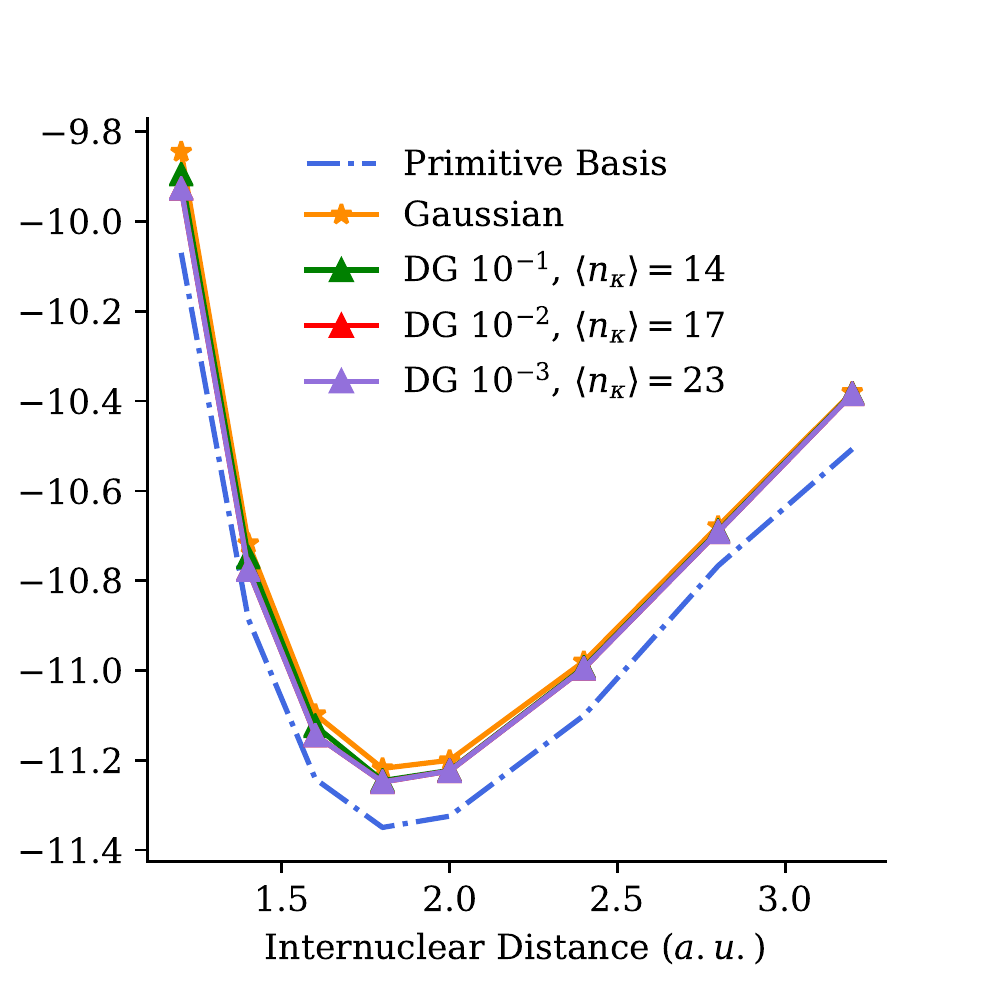}
    \caption{Potential energy surfaces for H$_{20}$ in DG basis with
    different tolerances and full plane wave basis. At equilibrium
    position, the average number of DG basis per atom is given by
    $\langle n_\kappa \rangle$ for each fit tolerance specified by DG
    $\epsilon$, where $\epsilon$ is the SVD cutoff.  For comparison the
    number of primitive functions per atom here is approximately $3000$.
    The primitive basis set is more expressive by design than the active
    space Gaussian cc-pVDZ basis against which the DG fit is performed.
    This allows even fairly loose DG fits to match the accuracy of the
    active space basis.}
    \label{fig:H20_kssolv}
\end{figure}

\begin{figure}[t!]
    \centering
    \includegraphics[width=0.40\textwidth]{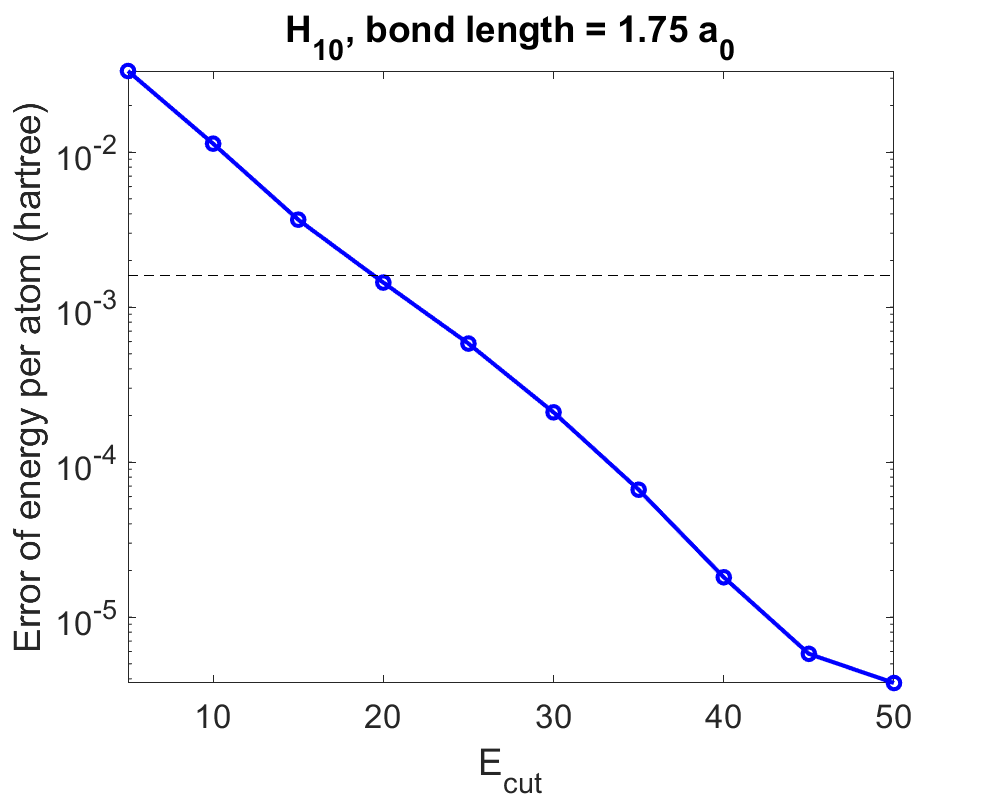}
    \caption{Convergence of the total energy per atom for a $H_{10}$ system with respect to the kinetic energy cutoff using the plane wave basis set and the ONCV pseudopotential. Chemical accuracy (black dashed line) is achieved around $E_{\text{cut}}=20$ hartree, which is the value used for all examples in this section.}
    \label{fig:ecut_h10}
\end{figure}

In order to demonstrate the performance of the DG approach for a long hydrogen chain,
we use the DFT-based MATLAB toolbox KSSOLV
\cite{yang2009kssolv}. KSSOLV uses a plane wave discretization in
combination with pseudopotentials to make larger systems computationally
tractable. Fig.~\ref{fig:ecut_h10} shows that 
chemical accuracy is achieved when the  kinetic energy cutoff is set to 
around $20$ hartree for a H$_{10}$ system using the ONCV pseudopotential. The kinetic energy cutoff needs to be much larger for all-electron calculations using the plane wave basis set.  With this kinetic energy cutoff, we can perform calculations on hydrogen chains
up to H$_{30}$. For H$_{30}$ the box size is $10\times 10\times 118 \, a_0^3$, and the grid size is $20\times 20\times 238$, or $95{,}200$ plane wave
dual basis functions (around $3{,}000$ basis functions per
atom).

We start by constructing an active basis set using the
cc-pVDZ basis with the given molecular geometry,  and
sample it with an underlying grid of plane wave dual functions. This
yields a real space discretization of the cc-pVDZ basis, which is then
transformed into a block-diagonal form by means of a DG blocking
procedure, also implemented in KSSOLV. As a technical note, the partitioning boundaries in the DG
approach are important. The hydrogen chains that are the subject of this
numerical investigation are quasi 1D-problems. This and the fact that we
use a real space discretization suggests that the ideal partitioning of
the basis in terms of a DG-scheme (see Section~\ref{sec:DG-basis})
corresponds to a non-uniform partitioning strategy, so that each
hydrogen atom is approximately located at the center of each element. We
remark that by construction the DG blocking procedure is able to produce
accurate results even if the partition is non-ideal, e.g. when an
hydrogen atom is located near the boundary of an element. However, this
can require a larger number of basis functions per atom compared to the
ideal partitioning strategy.

\begin{figure}[t!]
    \centering
    \includegraphics[width=0.40\textwidth]{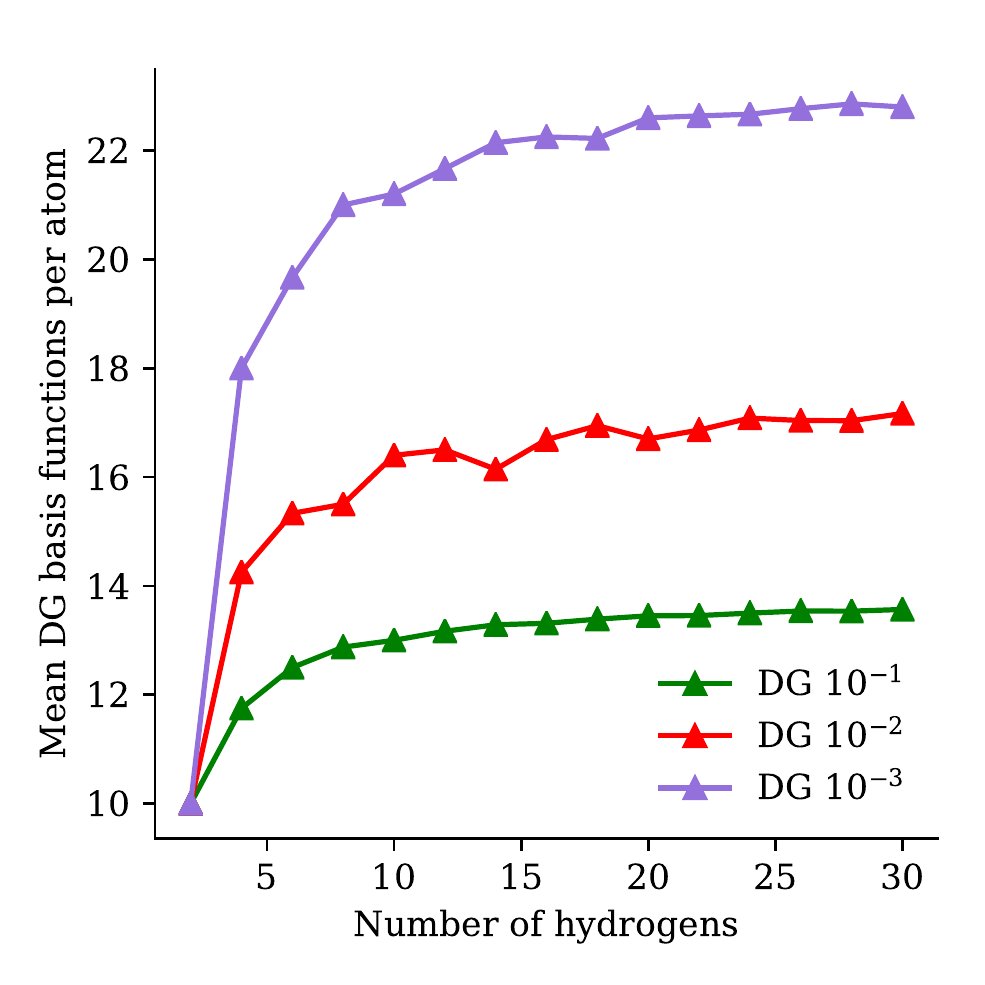}
    \caption{Convergence of block size $n_\kappa$ as a function of system size.  The average number of DG-basis per atom at equilibrium with SVD tolerances of $10^{-1}, 10^{-2}$ and $10^{-3}$.  For a fixed accuracy, it is observed that the average number of DG functions per block converges as a function of system size.}
    \label{fig:AverageNumberDG_kssolv}
\end{figure}

We first verify the accuracy of our DG-basis in
Figure~\ref{fig:H20_kssolv} by comparing potential energy surfaces (PES)
in DG discretizations of different tolerances with the corresponding PES
in the primitive basis. The primitive basis has been
chosen to be much more expressive than the Gaussian active space basis,
showing a lower absolute energy.  This allows even a coarse DG fit to reliably
match the Gaussian active space accuracy, while being far more compact
than the original primitive basis set. In fact, the energies obtained from the DG basis are slightly lower than those from the Gaussian basis set. This is because as $n_{\kappa}$ increases, the span of the Gaussian basis set becomes approximately a subspace of the span of the DG basis set, and hence the DG basis can possibly yield lower energies due to the variational principle. Also due to the variational principle, the energies obtained from the DG basis are noticeably higher than those from the primitive plane wave basis set, and the main limiting factor is the cc-pVDZ basis set to which DG is fitted.  The average number of
functions per atom $\langle n_\kappa \rangle$ is shown for each
tolerance and is approximately $20$, as compared to the primitive basis
which is built from roughly $3{,}000$ functions per atom.  
The results suggest
that the overall energetic accuracy is relatively insensitive even to
rather aggressive singular value cutoffs in the DG blocking procedure.

\begin{figure}[t!]
    \centering
    \includegraphics[width=0.40\textwidth]{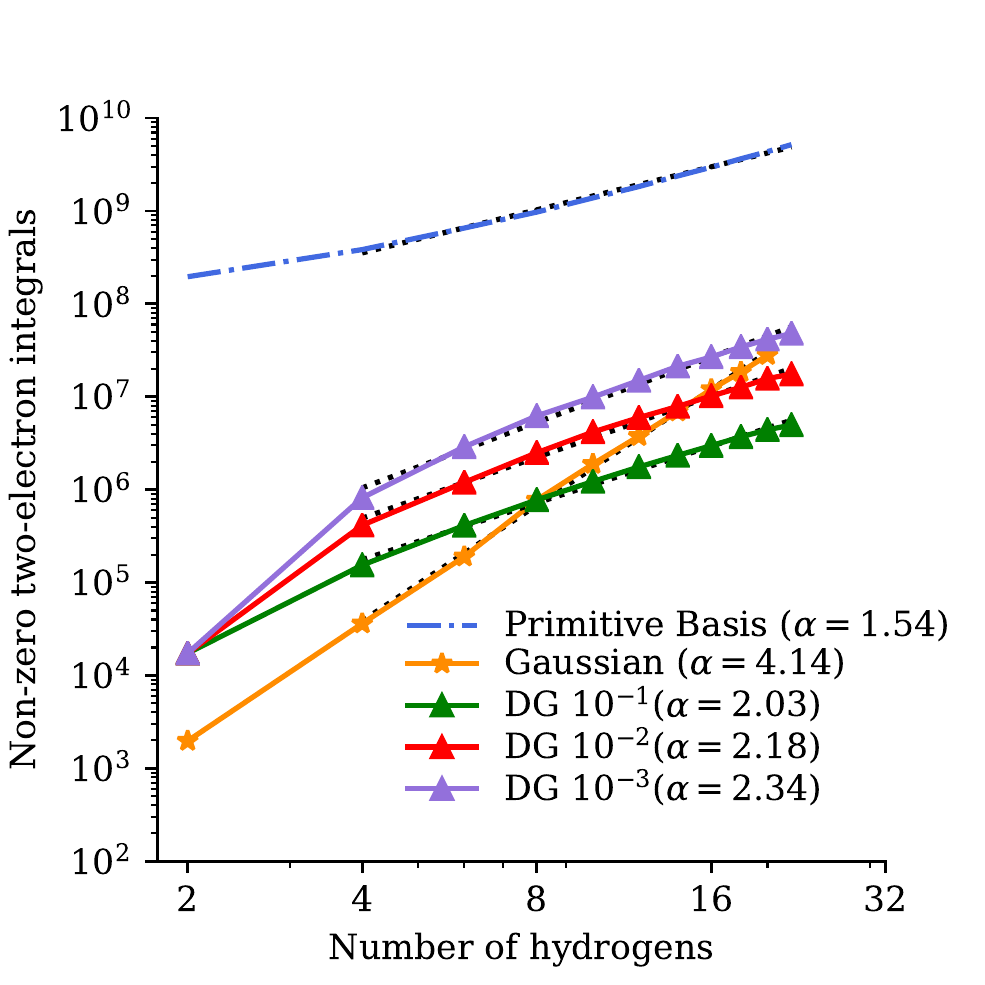}
    \caption{The number of non-zero two-electron integrals in different representations, for equilibrium and dissociation bond lengths with SVD truncation tolerances of $10^{-1}, 10^{-2}$ and $10^{-3}$, plotted on a log-log scale. We fit a trendline plotted with black dots from the second point onward to extract the scaling as a function of system size as $N^{\alpha} + c$ for some constant $c$, and list the exponent $\alpha$ beside each representation in the legend.  As predicted, for these system sizes the number of two-electron integrals lies between the primitive and active space representations, tending closer to the $O(N^2)$ scaling of the primitive representation, requiring fewer functions.  Note that for the Gaussian basis set, certain elements of the two-electron integrals can vanish due to the symmetry of the atomic configuration of the hydrogen chain. This has a larger impact for small system than for large systems, and therefore the scaling is observed to be slightly larger than $O(N_a^4)$.}
    \label{fig:Nonzero2EI_kssolv}
\end{figure}

We then plot the average number of DG-basis functions per atom in
Figure~\ref{fig:AverageNumberDG_kssolv} for fixed SVD thresholds in the
DG blocking procedure.  The data shows that as system size grows,
$\langle n_\kappa\rangle$ approaches a constant as a function of system
size for a fixed tolerance as reasoned earlier.  The combination of the
energetic insensitivity to cutoff and the approach to a constant
$\langle n_\kappa\rangle$ strongly supports the existence of a crossover
regime where DG is more cost effective than other representations. We
confirm the crossover directly by examining the quantities most relevant
for fault-tolerant cost models of chemistry on a quantum computer, in
particular the number of non-zero two-electron integrals in each
representation as well as the $\lambda$ factor Eq.~\eqref{eqn:lambda_dg}
in Figures~\ref{fig:Nonzero2EI_kssolv} and \ref{fig:Lambda_kssolv},
respectively (for $\lambda$ factors for different bond lengths see
Appendix~\ref{App:DG-PW}, Figure~\ref{fig:Nonzero2EI_full_kssolv}).  A
cutoff of $10^{-6}$ is used to count an individual integral as $0$ when
calculating these quantities empirically on the systems of interest.  We
choose to illustrate the cost crossover for a bond length of $1.7a_0$
since we here detected the largest deviation with respect to the
truncation tolerance. 

Considering the cost-model in the fault-tolerant setting outlined in
Section \ref{sec:LCU_cost}, one can already observe a scaling advantage
for the DG representation over simple Gaussian based active space
representations. Recall from that section that the cost using an LCU
method to evolve for some time $t$ is roughly $O(\sqrt{L} \lambda t)$
where $L$ is the number of non-zero terms in the Hamiltonian.  We
consider here only the two-electron integrals as they represent the
dominant cost contribution in most cases.  For molecular orbital
representations, this is generally $O(N_a^2 \lambda t)$, which was
subsequently improved to $O(N_a^{3/2} \lambda t)$ using low-rank
structure in the problem. 
For the hydrogen chains examined here in the Gaussian basis set, we see
empirically over the system sizes considered $L \propto N_h^{4}$
 and $\lambda \propto N_h^{2.5}$, where
$N_h$ is the number of hydrogen atoms in the chain, leading to an
expected cost scaling of $O(N_h^{4.5} t)$ when not exploiting further
low rank structure.  In the DG representation for the same problem, we
see that $L \propto 2.25$ and $\lambda \propto 1.5$, which suggests an
empirical scaling for this physical system of $O(N_h^{2.6}t)$.  The same
calculation for the primitive basis suggests a scaling of $O(N_h^{1.8})$,
which is the lowest seen, however, the simulation cost and qubit counts
required are many orders of magnitude higher here as seen in the
figures.  In both quantities we observe a constant factor crossover
where the DG representation has strictly lower unique two-electron
integrals and $\lambda$ factor when compared to the Gaussian active
space basis at around 15 to 20 hydrogen atoms.  This suggests that the
DG representation is not only advantageous in a scaling sense, but also for modest finite system sizes for fault-tolerant implementations.  

\begin{figure}[t!]
    \centering
    \includegraphics[width=0.40\textwidth]{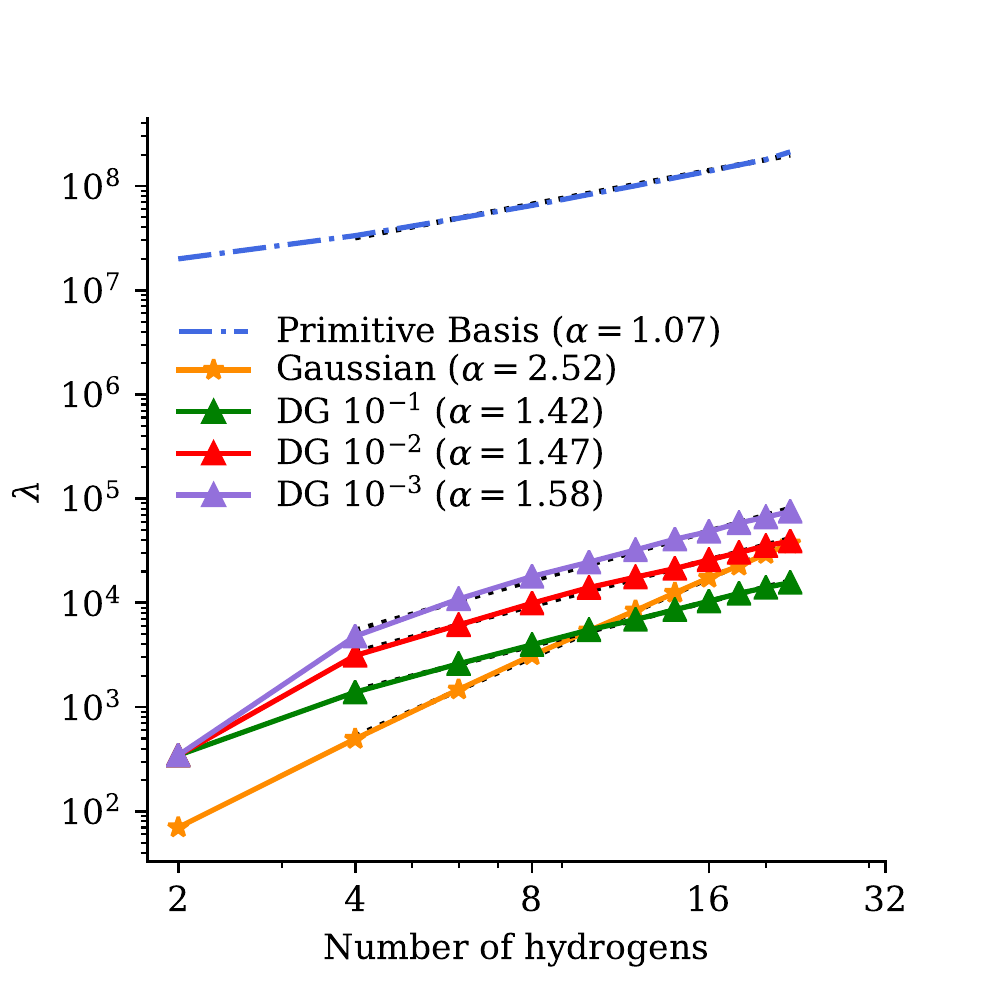}
    \caption{$\lambda$ value for Gaussian and DG basis. A core quantity in determining the cost in quantum algorithms, $\lambda$, is plotted as a function of system size for different representations. The notation DG $\epsilon$ indicates an SVD cutoff of $\epsilon$ in the blocking procedure. We observe an advantageous crossover before or around H$_{20}$ in all cases with respect to an actual value.  We fit a trendline plotted with black dots from the second point onward to extract the scaling as a function of system size as $N^{\alpha} + c$ for some constant $c$, and list the exponent $\alpha$ beside each representation in the figure legend. We see the scaling for the DG basis is significantly better in all cases than for the active space basis as well.}
    \label{fig:Lambda_kssolv}
\end{figure}

\subsection{Correlated calculations in a DG Gausslet basis}
Here we demonstrate that the performance of the DG basis set for a vastly different regime, which uses a natural orbital active space from an exact DMRG calculation fit to Gausslet primitive functions in a correlated calculation.  The Gausslet basis set is a recent approach to improve the discretization of quantum chemical problems\cite{white2017hybrid,white2019multisliced}. It has a special focus on sparsity, spatial locality, and orthonormality, to fulfill the needs of strong-correlation methods like the DMRG, while keeping the number of basis functions lower than other grid based bases. Our calculations illustrate the accuracy of the CCSD method in a DG-basis with an underlying primitive Gausslet basis set with respect to the original active space. The block-diagonal form of the Hamiltonian in a DG-basis yields an asymptotic improvement in the number of non-zero two-electron integrals, which is directly related to the circuit depth and size. The calculations presented here are limited to short hydrogen chains due to the size of the Gausslet basis set and expense of correlated calculations.  The goal of these calculations is to demonstrate that the DG approach maintains accuracy for correlated calculations due to its construction as a fit to the span of the active basis set, rather than a density matrix generated with a fixed level of theory.

\begin{figure}[t!]
    \centering
    \includegraphics[width=0.40\textwidth]{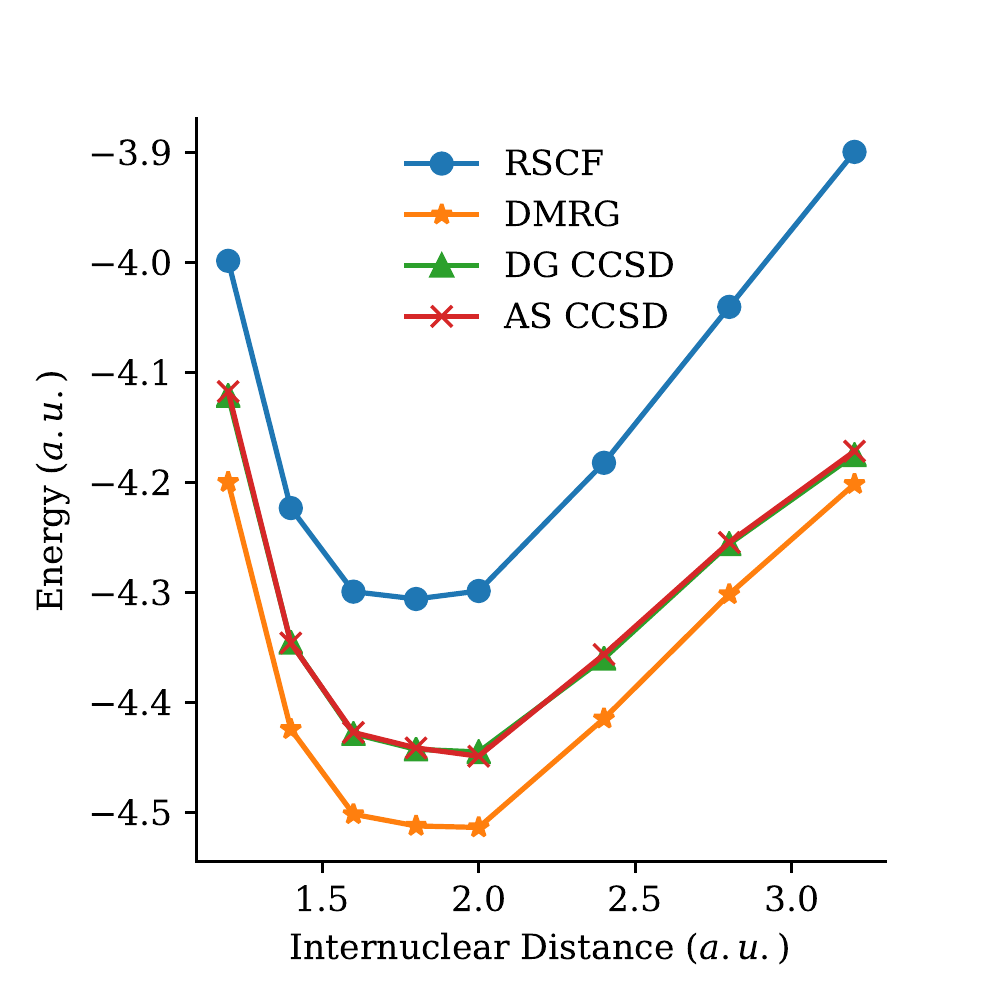}
    \caption{Correlated potential energy surfaces for H$_8$.  We plot the potential energy surface computed using the Gausslet basis as the primitive basis and a subset of the exact DMRG natural orbitals as an active space (AS).  The CCSD energy is calculated in the active space as well as the DG fit to that active space, showing excellent agreement with the active space energetics in a correlated calculation.  The exact DMRG energy in the more expressive, full primitive basis (Gausslets) is shown for reference.}
    \label{fig:H8_Gausslet_PES}
\end{figure}

Specifically, we compare CCSD energy results in an active space basis set with calculations in a DG-basis fit to the same active space basis.  This serves to quantify the overhead in restricting the basis to have block locality versus the totally delocalized natural orbital active space.  The calculations are performed for H$_2$, H$_4$, H$_6$ and H$_8$ with varying symmetrically stretched bond lengths. Figure~\ref{fig:H8_Gausslet_PES} shows the calculation for H$_8$ with optimal DG partitioning of the Gausslet basis set, the respective figures for H$_2$, H$_4$, and H$_6$ are presented in Appendix~\ref{App:Corr_calc} (Figures~\ref{fig:H2_Gausslet_PES},~\ref{fig:H4_Gausslet_PES},~\ref{fig:H6_Gausslet_PES}). From these plots one may see that enforcing a block-diagonal structure of the Hamiltonian does not reduce the accuracy further from the active space approximation.
 
The Gausslet basis for computations resulting in Figure~\ref{fig:H8_Gausslet_PES} consists of $1{,}336$ Gausslets corresponding to an average of 167 functions per atom. This is far lower than that in the plane wave dual basis reflecting the variable resolution available with the Gausslet basis. 
Averaging over all values along the PES, we find that the full two-electron integral tensor for this basis has $\sim 2{,}310{,}318$ non-zero elements. Using a DG basis it is possible to reduce this number to $\sim 511{,}449$ non-zero two-electron integrals without losing accuracy compared to the respective active space approximation. This suggests cost reductions for correlated calculations that are similar to the observations in Section~\ref{sec:DG-DFT-Calc}. Note, however, that H$_8$ is a small system compared to the systems considered in Section~\ref{sec:DG-DFT-Calc}, consequently, the number of non-zero two-electron integrals in the DG-basis is still larger than in the active space approximation ($\sim 52{,}346$). This aligns with the results from the previous section, which showed that the improvement of the non-zero two-electron integral count for the DG-basis becomes observable for 15 to 20 atoms, depending on the imposed truncation tolerance. A naive extrapolation of the non-zero two-electron integral count in Table~\ref{tab:numerical_values_Gausslets} (see Appendix~\ref{App:Corr_calc} Figure~\ref{fig:nnz_tei_corr_calc_extrapolation}) suggests a crossover around 25 atoms for the coupled-cluster calculations, which is again in agreement with the computations in Section~\ref{sec:DG-DFT-Calc} with a low truncation tolerance (see Appendix~\ref{App:DG-PW} Table~\ref{tab:numerical_values_crossover}). We conclude that the trial calculations for small systems performed here together with results from Section~\ref{sec:DG-DFT-Calc} suggest that the DG approach maintains the accuracy of active space approximations for correlation methods but with a more efficient representation on quantum devices once a certain system size is reached.

\subsection{DMRG calculations in a DG basis}
Here we examine the power of the DG technique to compactly and cheaply represent problems for classical, correlated DMRG calculation by constructing an active space that takes static correlation from a UHF calculation done in a flexible primitive basis set such as Gausslets, combined with contributions from a basis set that has been empirically refined to capture dynamic correlations. These calculations have the dual purpose of demonstrating the power of the DG approach even in highly correlated calculations more generally, including calculations on a quantum computer.  Here we demonstrate a hybrid active space approach for cheaply finding a balanced active space.  Specifically, after performing a UHF calculation in a Gausslet basis, we project both the cc-pVDZ basis set and the UHF orbitals onto the primitive Gausslet basis. The UHF calculation can be performed cheaply.
By including the UHF orbitals in our active space, we ensure that there are no HF-level errors in the basis. The only lack of completeness is associated with correlation beyond HF. To (partially) capture correlation in a compact way, we include contributions from the empirical Gaussian basis with a slightly smaller factor, $\alpha=0.01$ here, then perform the DG blocking procedure to develop the basis as before.  The resulting basis maintains the flexibility for the HF solution in the Gausslet basis for the low lying orbitals, while now including the refined features of the Gaussian orbitals without losing the block diagonal structure.   Although it is difficult to compare precisely of the efficiency of the primitive and DG bases, roughly we find that this approach can achieve a 1-2 orders of magnitude reduction of computational cost over either the primitive Gausslet basis or Gaussian basis set, while achieving excellent accuracy with respect to the complete basis set limit.

\begin{figure}[t!]
    \centering
    \includegraphics[width=0.40\textwidth]{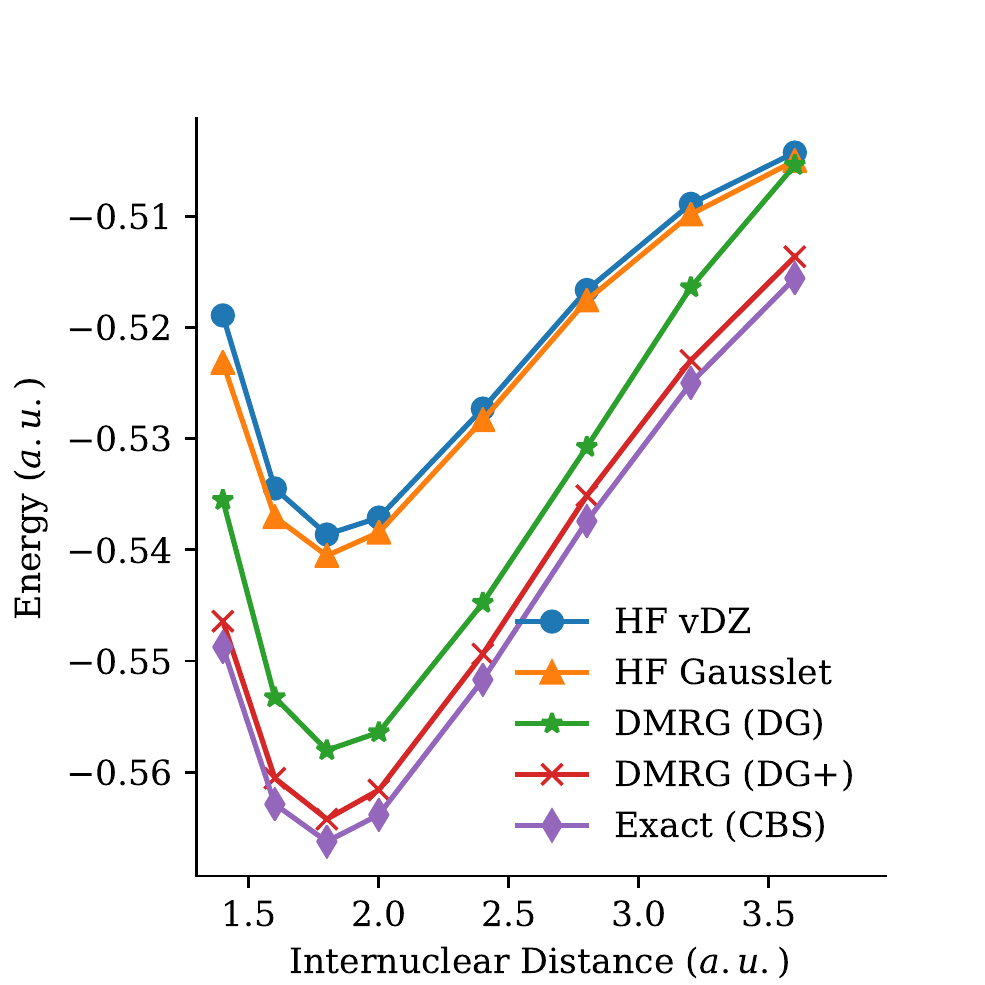}
    \caption{Potential energy surfaces for H$_{10}$ constructed with the hybrid approach.  We plot the potential energy surface for unrestricted Hartree--Fock (HF) calculations in both a cc-pVDZ (vDZ) Gaussian basis and a Gausslet basis, from which the DG basis sets are derived to perform a DMRG calculation.  We then contrast this with a calculation done in the DG basis based using the first 7 virtuals from the HF-DG basis (DG), and finally showcase the power of the hybrid approach to cost-effectively re-introduce dynamic correlation without the cost of the original Gaussian basis by adding a point sampling of the cc-pVDZ basis into the DG basis (DG+).  This hybrid basis attains nearly the exact solution with respect to the complete basis set (CBS) limit with a fraction of the cost of using either the Gausslet or Gaussian basis set directly.}
    \label{fig:H10_DMRG_Gausslet_PES}
\end{figure}

Numerical calculations for the $H_{10}$ system are shown Fig. \ref{fig:H10_DMRG_Gausslet_PES}.  For comparison, we perform an unrestricted Hartree-Fock calculation in both a traditional Gaussian basis set, cc-pVDZ, and a multi-sliced Gausslet basis set.  The Gaussian basis set contains 5 spatial orbitals per atom, totaling to 50 spatial orbitals with the associated non-diagonal two-electron integrals as one would expect.  The Gausslet basis is formed adaptively according to pre-determined cutoffs, and the number of functions ranges from 7000-10000 spatial orbitals for these calculations, while retaining the diagonal property, making calculations at the UHF level relatively straightforward, even with such a formidable number of basis functions.  This large primitive basis gives
UHF results near the complete basis set limit, well beyond the accuracy of this Gaussian basis. This high accuracy in the HF comes at little cost;
the UHF is still fast compared to the correlated calculations and the large number of primitive functions do not strongly affect the
size of the DG basis. For larger systems the accuracy could be reduced to keep the HF manageable.

To construct the DG basis, we construct 10 spatial blocks, 1 around each atom.  We make use of the UHF calculation density matrix and keep 7 total orbitals per block, yielding 70 total spatial orbitals, a number almost identical to the number of Gaussians in the cc-pVDZ basis, but maintaining the block diagonal property of the two-electron integrals.  By construction, UHF in this basis can accurately match the UHF results of the Gausslet basis set.  The introduction of correlation through DMRG on this basis shows improvement as expected but a relative offset from the exact answer due to UHF's focus on static correlation.  By using the weighting procedure to the include in the DG construction some of the cc-pVDZ weight for dynamic correlation, at a weighting factor of $\alpha=0.01$, we find excellent agreement with calculations done in the exact basis set limit.  The number of functions kept here is 15 per block, yielding 150 functions total, or about 3 times that of a cc-pVDZ basis.  However, the structure of the interactions and spatially local construction of the DG functions allows the DMRG calculation to be done with a 1-2 order of magnitude reduction in computational cost as compared to using the DG or Gaussian basis sets alone for the current implementation.  This suggests the hybrid active space approach with the DG blocking procedure is a powerful technique for recovering both static and dynamic correlation in a cost effective manner.

To elaborate on the scaling of DMRG, $m$ be the bond dimension for the state required for the desired accuracy.
Then the computational cost of the Gaussian basis sets is expected to be $O(N_a^3 m^3)$.  In contrast, for Gausslets and DG representations based on Gausslets, using matrix product operator (MPO) compression and let $D$ be the MPO dimension, one expects the asymptotic cost to be $O(N_p D m^3)$ and $O(N_d D m^3)$, respectively.   
Due to the localization properties, $D$ is expected to depend weakly on length, and be comparable for both the Gausslet and DG block representations.  One can see this from considering the MPO decomposition in the Gausslet representation then transformed to the DG representation.  While some expansion of the bond dimension could happen within a block, it is bounded by the Gausslet dimension of the block from the properties of a Schmidt decomposition, and no inter-block mixing occurs.  As a result, the bond dimension will be comparable.  Hence, it is the massive reduction in number of basis functions, from 10,000 to 150 that reduces the cost by several orders of magnitude while maintaining excellent accuracy.  The Gaussian basis set, while advantageous in number of functions, suffers from lack of spatial locality, but the cost of this non-locality in terms of the required bond dimension for equivalent accuracy has not been studied, and likely depends significantly on the completeness of the basis. It is
easier to compare the scaling of the Gaussian DMRG due to the lack of diagonality, resulting in a much larger set of two-electron integrals. In this case, the block diagonality of the DG basis typically results in approximate linear scaling in the number of atoms, versus cubic dependence with Gaussian DMRG. Hence in both cases, the DG approach offers a significant reduction in computational cost for an accuracy that nearly approaches the complete basis set limit, which was obtained through accurate quantum Monte Carlo calculations~\cite{Motta:2017,white2019multisliced}.

\section{Conclusion}
The discretization problem is a crucial aspect determining the cost and effectiveness of quantum chemistry methods both for classical and quantum methods.  The most popular basis sets for correlated calculations, Gaussian and molecular orbital representations, are notably more compact than alternatives and have a well develop set of tools for their use.  Unfortunately, in cost models for quantum computation, the overhead of a quartic number of terms in the Hamiltonian leads to poor scaling with system size.  On the other side of the spectrum, representations that achieve a strictly quadratic number of terms and a diagonal representation, such as Gausslets or plane wave dual functions exhibit excellent scaling with system size, but have overheads that make them undesirable for modest size implementations.

Here we introduced a systematic method for interpolating between the two regimes through the use of a blocking procedure, motivated from the discontinuous Galerkin (DG) method.  This method is able to use any primitive basis with the diagonal property to represent a delocalized active space basis, while maintaining the diagonal property between blocks.  By choosing a plane wave dual primitive basis and Gaussian active space, we were able to show how one can adaptively interpolate between these two regimes to attain both a scaling and constant factor advantage over the target active space.

When these empirical results are put into the context of known costs for exact quantum algorithms for chemistry, we observed a scaling improvement over Gaussian basis sets from $O(N_h^{4.5})$ to $O(N_h^{2.6})$ with a constant factor crossover around 15 to 20 hydrogen atoms.  This suggests that for modest sized systems, such as those just beyond the classically tractable regime, this representation will be the optimal choice for quantum algorithms.  Moreover, we showed that for high accuracy DMRG calculations, one may take advantage of this representation to achieve a high accuracy calculation with a cost reduction that is over an order of magnitude with respect to traditional representations.  In all cases, one may use this methodology to scale between a compact representation and one with superior integral scaling depending on the requirements of a particular method.

\subsection*{Acknowledgments} 

This work was partially supported by 
the Department of Energy under Grant No. DE-SC0017867,
the Quantum Algorithm Teams Program under Grant No. DE-AC02-05CH11231,
the Google Quantum Research Award (L.L.), 
the Research Council of Norway under CoE Grant No. 262695, 
the Peder Sather Grant Program (F.M.F), 
the NASA Space Technology Research Fellowship (B.O.),
and the Ning fellowship (Q.Z.).  

\bibliographystyle{apsrev4-1_with_title}
\bibliography{references}

\appendix
\section{Trotter step by low-rank factorization}
\label{app:low_rank}

As an alternative to the fixed swap networks used in the main text to evolve under the two-electron integral term in the DG basis, one may use the low-rank factorization  
strategy in~\cite{Motta2018}, but applied to the block diagonal matricized tensor $v_{\kappa, \kappa', \lambda, \lambda' ; i,i',j,j'}^{(d)}$.  For such a matrix, we know the maximum number of Cholesky factors is given by the dimension of the matrix, $N_d^2$, however, because of the block structure, each of the Cholesky factors need only have non-trivial support only within a $\kappa, \kappa'$ block.  If $n_\kappa$ bounds the larger of $n_\kappa$, $n_{\kappa'}$, then the dimension of one of these blocks is $O(n_\kappa^2)$.  It's easy to see that the total matricized tensor has dimension $O(n_\kappa^2 N_b^2)$, and a number of non-zero entries scaling as $O(n_\kappa^4 N_b^2)$.  Hence it matches the primitive limit as $n_\kappa \rightarrow 1$ and $N_b \rightarrow N_d$ and the active space limit as $n_\kappa \rightarrow N_d$ and $N_b \rightarrow 1$.

To execute a single Trotter step of the two-electron part of the Hamiltonian, one may start from a factorization that is a product over $\kappa, \kappa'$ blocks as
\begin{align}
    & \exp \left[ -i  \Delta t \sum_{\kappa,\kappa';i,i',j,j'}
  v^{(d)}_{\kappa,\kappa';i,i',j,j'} 
  \hat{c}^{\dagger}_{\kappa,i} \hat{c}^{\dagger}_{\kappa',i'}
  \hat{c}_{\kappa',j'} \hat{c}_{\kappa,j} \right]\approx \notag \\
    & \prod_{\kappa, \kappa'} \exp \left[ -i  \Delta t \sum_{i, i', j, j'} v^{(d)}_{\kappa,\kappa';i,i',j,j'}
  \hat{c}^{\dagger}_{\kappa,i} \hat{c}^{\dagger}_{\kappa',i'}
  \hat{c}_{\kappa',j'} \hat{c}_{\kappa,j}\right] \notag \\
   & \equiv \prod_{\kappa \kappa'} R_{\kappa \kappa'}
\end{align}

Within a non-trivial block $(v_{\kappa, \kappa', \lambda, \lambda' ; i,i',j',j'}^{(d)} = v_{\kappa, \kappa'; i,i',j',j'}^{(d)})$ we expect the following decomposition
\begin{align}
  v_{\kappa,\kappa' ; i,i',j',j'}^{(d)} \approx \sum_{\mu\nu\ell} \Big(& U^{(\ell,\kappa, \kappa')}_{i\mu} U^{(\ell,\kappa, \kappa')}_{i'\mu} 
U^{(\ell,\kappa,\kappa')}_{j\nu}\nonumber\\
& \cdot U^{(\ell,\kappa, \kappa')}_{j'\nu} \lambda^{(\ell,\kappa, \kappa')}_{\mu} \lambda^{(\ell,\kappa, \kappa')}_{\nu}\Big). 
  \label{}
\end{align}
Here $\ell$ comes from Cholesky factorization, and $\mu,\nu$ comes from
a second eigenvalue decomposition.  As for a single block, $\kappa, \kappa'$, we assume $n_{\kappa}$ is independent of the system size, so are the index ranges of
$\mu,\nu,\ell$. For each $\ell,\kappa,\kappa'$,
$U^{(\ell,\kappa,\kappa')}$ is a matrix with orthogonal columns.  This decomposition allows us to apply $R_{\kappa \kappa'}$ using the low-rank decomposition technique.  The maximum rank of these factors is $L$, which empirically for Gaussian basis sets we expect to scale as $O(n_\kappa)$, but due to the lack of empirical data for large DG basis sets we assume the worst case, $O(n_\kappa^2)$.  For each of these factors, the second eigenvalue decomposition will have maximal rank $\rho_{\ell} = O(n_\kappa)$.  The depth of such circuits using a fermionic swap network scales as $\sum_{\ell}\rho_{\ell} < L n_\kappa$, which in empirical studies on molecular orbital basis sets scaled at least as $\Omega(n_\kappa \log n_\kappa)$ and in the worst case $O(n_\kappa^3)$. In any case, for a given error tolerance, from locality we know that $n_\kappa$ converges to a constant as a function of system size, and hence each $R_{\kappa \kappa'}$ may be executed in constant depth for large system using the low-rank Trotter step, which simultaneously executes a fermionic swap network.

Following this idea further, if one constructs a system mapped to qubits in 1D, where the mapping to orbitals is grouped by $\kappa$, then the $\kappa \kappa'$ block may be executed for $\kappa'= \kappa+1$, and at the same time, the two blocks may be swapped with this technique.  Moreover, as this adjacent $\kappa \kappa'$ block is disjoint from the qubits outside of $\kappa \kappa'$ this operation may be parallelized across pairs of adjacent blocks.  This corresponds exactly to lifting the original fermionic swap network for time evolution to the level of blocks, where the individual fermionic simulation operations (as they were referred to in the original work) are performed via the low-rank method.  From this, we see that the total required depth scales as $O(N_b)$ as desired.   Although we have assumed here that $n_\kappa$ approaches a constant as system size grows, if we account for this in the worst case, we expect a depth of $O(N_b n_\kappa^3)$, which matches the fixed swap networks in the main body of the text.  Hence, determining the optimal choice of implementation will be done to constant factors and requires further empirical study.

\section{Scaling crossover in a DG plane wave dual basis}
\label{App:DG-PW}

\begin{figure}[t!]
    \centering
    \includegraphics[width=0.50\textwidth]{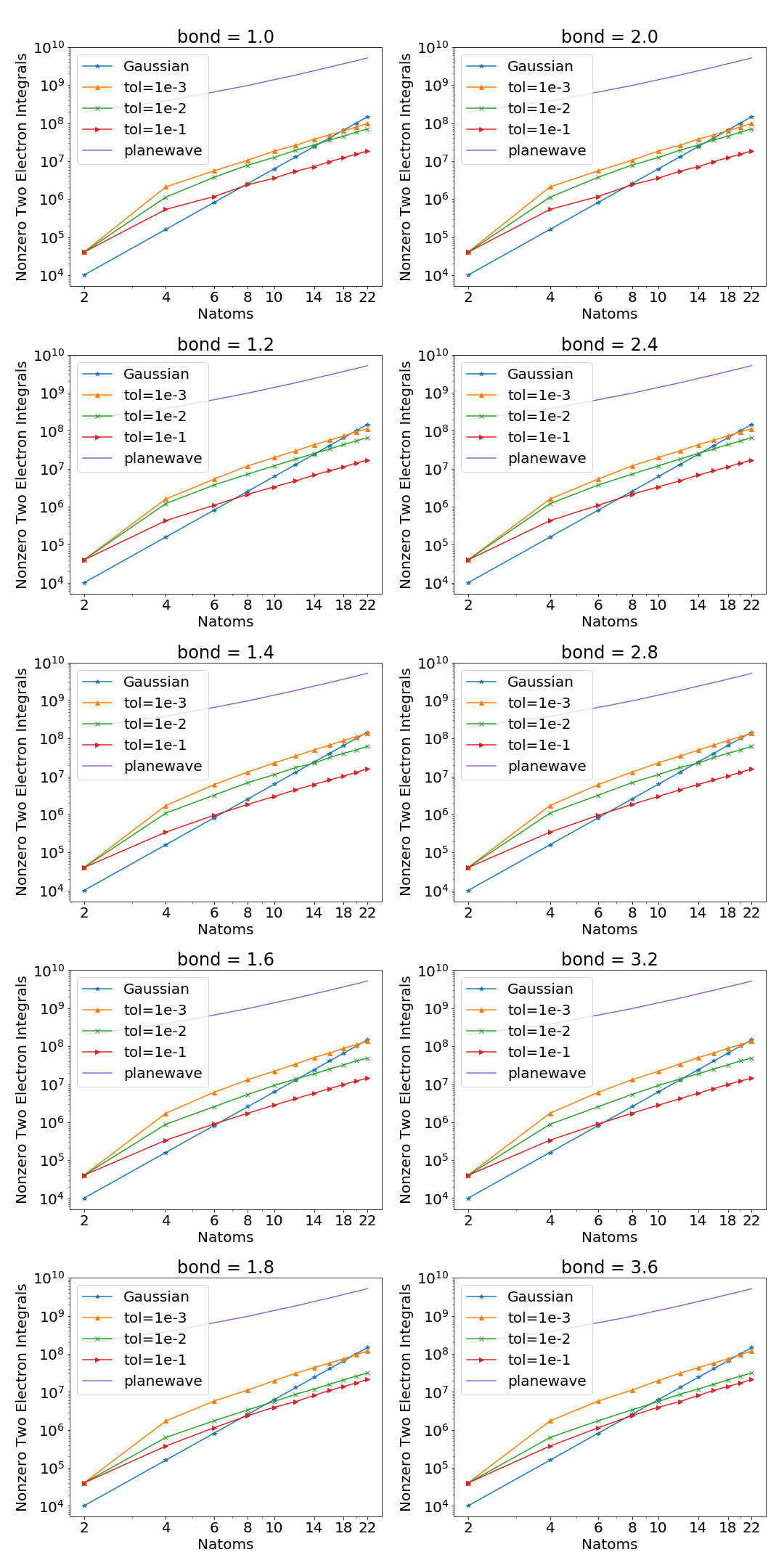}
    \caption{The number of non-zero two-electron integrals with bond lengths $= 1.0$ to $3.6$ in atomic units, tol $= 10^{-1}$ to $10^{-3}$, plotted on log-log scale. The crossover appears before or around H$_{22}$ in all cases.
    \label{fig:Nonzero2EI_full_kssolv}}
\end{figure}

Besides the total number of two-electron integrals, the cost of certain quantum
algorithms such as the LCU method depends also on $\lambda$, the sum of the absolute value
of the two-electron integrals.  Here $\lambda$ is computed as 
\begin{equation}
\label{eqn:lambda_dg}
    \lambda = \sum_{\{\kappa,\kappa';pqrs\}/\{p=r,q=s\}}|v_{\kappa,\kappa';p,q,r,s}|.
\end{equation}
Notice that the terms $\{p=r,q=s\}$ nominally diverge in a plane wave basis set with periodic boundary conditions. These values can be computed by including correction terms to the periodic boundary condition~\cite{GygiBaldereschi1986,BylaskaTsemekhmanBadenEtAl2011}. However, the number of such terms in the two-body interaction scales the same as that in the one-body interaction in the Hamiltonian, so we omit such low-order terms directly for simplicity.  In Figure~\ref{fig:Lambda_kssolv}, we observe that the crossover point occurs around H$_8$ when the tolerance is set to $0.1$, and around H$_{18}$ when the tolerance is $0.01$. 


\begin{table}[h!]
\centering
\begin{tabular}{c|l|l|l}
\toprule
Tolerance & $10^{-3}$ & $10^{-2}$ & $10^{-1}$ \\
\hline
$b=1.0$ &H$_{16}$-H$_{18}$ &H$_{14}$-H$_{16}$ &H$_8$-H$_{10}$\\
$b=1.2$ &H$_{18}$-H$_{20}$ &H$_{14}$-H$_{16}$ &H$_6$-H$_8$\\
$b=1.4$ &H$_{20}$-H$_{22}$ &H$_{14}$-H$_{16}$ &H$_6$-H$_8$\\
$b=1.6$ &H$_{20}$-H$_{22}$ &H$_{12}$-H$_{14}$ &H$_6$-H$_8$\\
$b=1.7$ & H$_{20}$-H$_{22}$ &H$_{10}$-H$_{12}$ &H$_6$-H$_8$\\
$b=1.8$ &H$_{18}$-H$_{20}$ &H$_8$-H$_{10}$ &H$_8$-H$_{10}$\\
$b=2.0$ &H$_{16}$-H$_{18}$ &H$_8$-H$_{10}$ &H$_8$-H$_{10}$\\
$b=2.4$ &H$_{10}$-H$_{12}$ &H$_8$-H$_{10}$ &H$_8$-H$_{10}$\\
$b=2.8$ &H$_{8}$-H$_{10}$ &H$_8$-H$_{10}$ &H$_8$-H$_{10}$\\
$b=3.0$ &H$_{8}$-H$_{10}$ &H$_8$-H$_{10}$ &H$_8$-H$_{10}$\\
$b=3.2$ &H$_{8}$-H$_{10}$ &H$_8$-H$_{10}$ &H$_6$-H$_8$\\
$b=3.6$ &H$_{8}$-H$_{10}$ &H$_8$-H$_{10}$ &H$_4$-H$_6$\\
\hline
\hline
\end{tabular}
\caption{Crossover regions for different bond lengths and different truncation tolerances.}
\label{tab:numerical_values_crossover}
\end{table}

\section{Correlated calculations in a DG Gausslet basis}
\label{App:Corr_calc}

\begin{figure}[t!]
    \centering
    \includegraphics[width=0.40\textwidth]{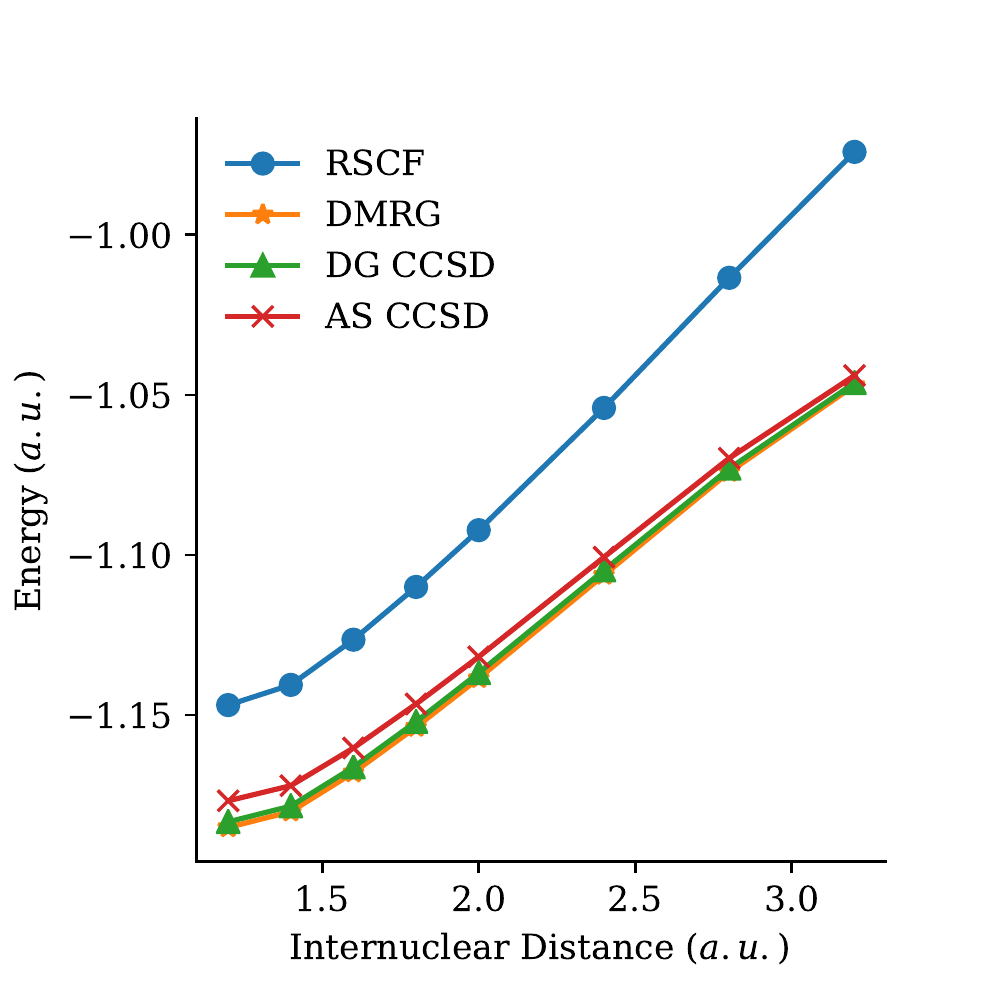}
    \caption{Potential energy surfaces for H$_2$ in a Gausslet basis computed by restricted SCF, (DG-)CCSD and benchmarked with DMRG results. The averaged number of non-zero two-electron integrals per atom (DG-element $\kappa$) is 4{,}144. The averaged number of non-zero two-electron integrals of the CAS calculations is 205.}
    \label{fig:H2_Gausslet_PES}
\end{figure}

\begin{figure}[t!]
    \centering
    \includegraphics[width=0.40\textwidth]{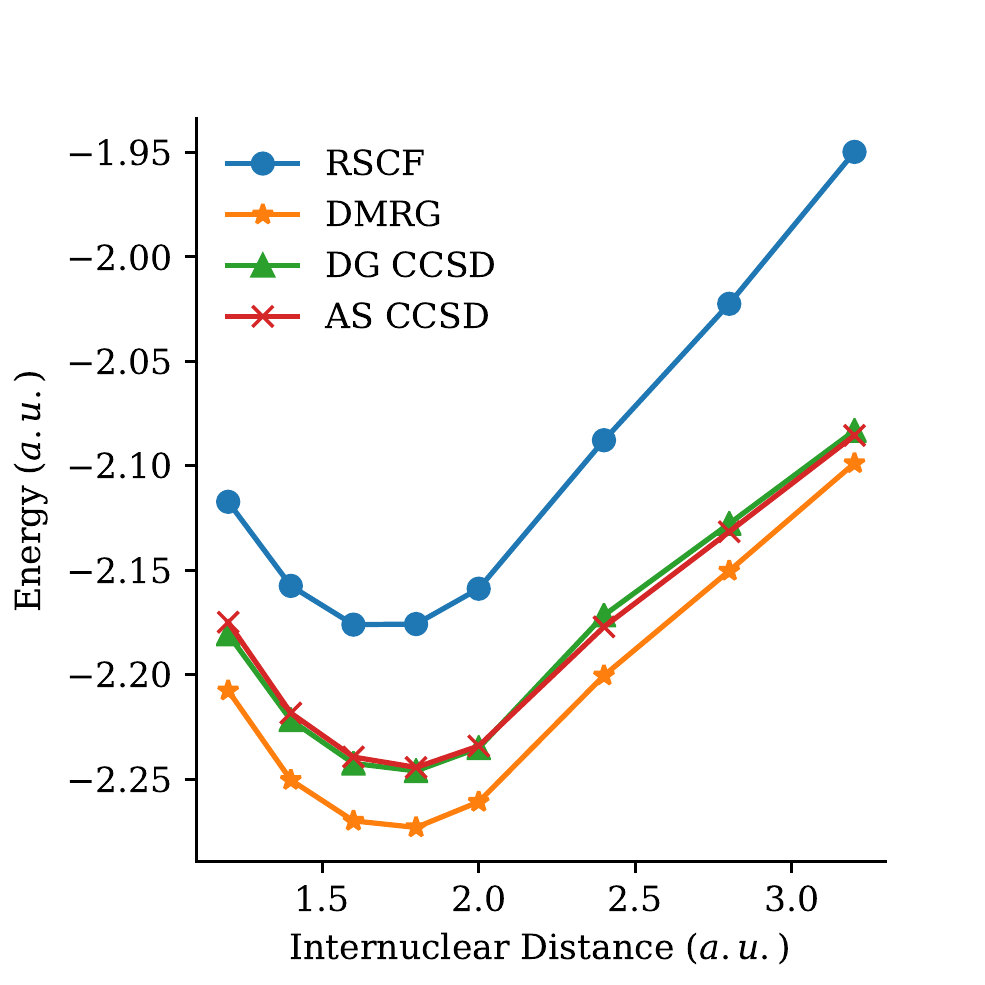}
    \caption{Potential energy surfaces for H$_4$ in a Gausslet basis computed by restricted SCF, (DG-)CCSD and benchmarked with DMRG results. The averaged number of non-zero two-electron integrals per atom (DG-element $\kappa$) is 16{,}580. The averaged number of non-zero two-electron integrals of the CAS calculations is 3{,}269.}
    \label{fig:H4_Gausslet_PES}
\end{figure}

\begin{figure}[t!]
    \centering
    \includegraphics[width=0.40\textwidth]{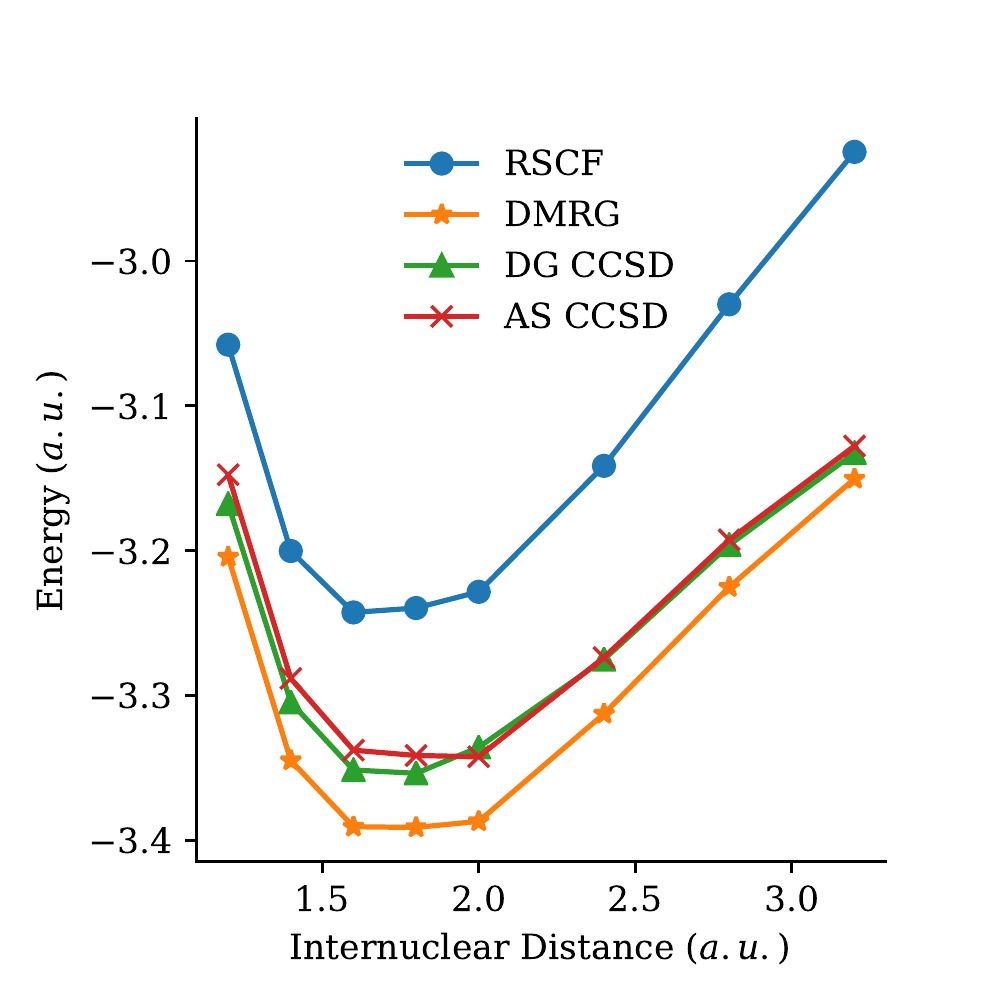}
    \caption{Potential energy surfaces for H$_6$ in a Gausslet basis computed by restricted SCF, (DG-)CCSD and benchmarked with DMRG results. The averaged number of non-zero two-electron integrals per atom (DG-element $\kappa$) is 287{,}762. The averaged number of non-zero two-electron integrals of the CAS calculations is 16{,}551.}
    \label{fig:H6_Gausslet_PES}
\end{figure}

\begin{figure}[t!]
    \centering
    \includegraphics[width=0.40\textwidth]{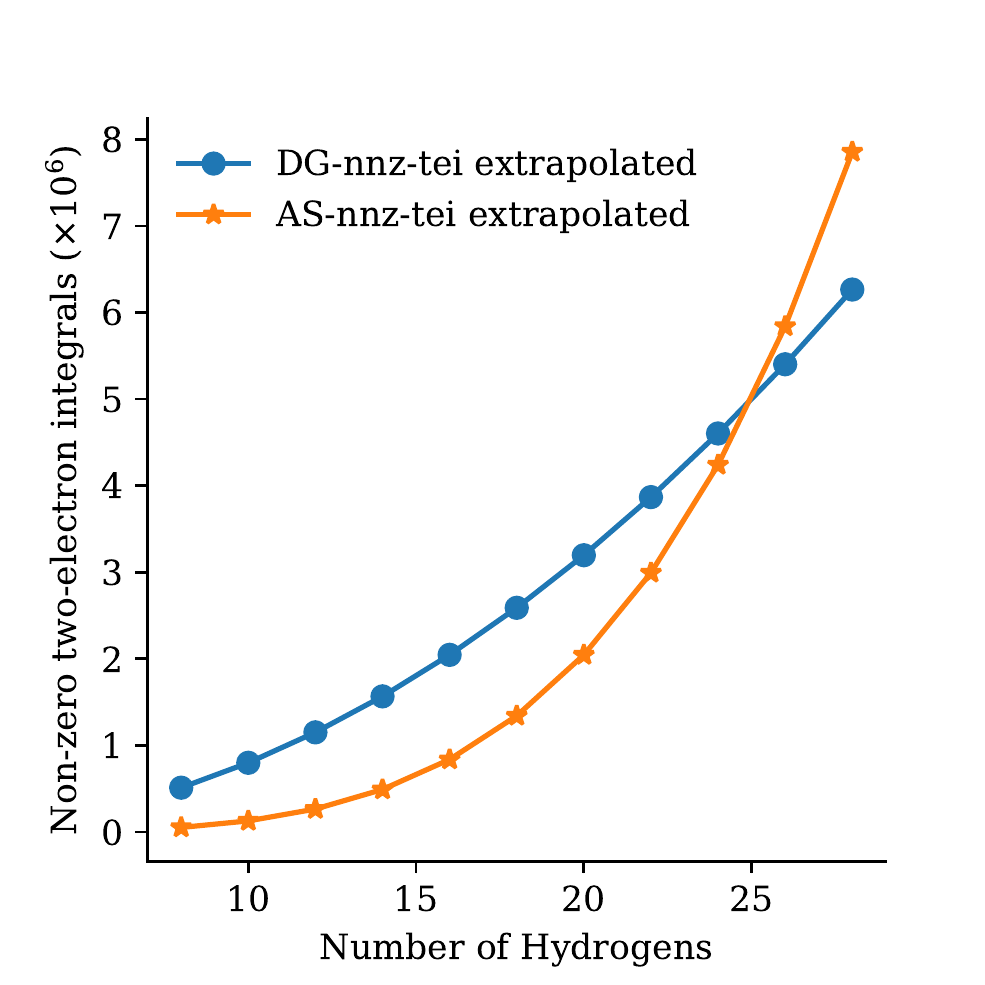}
    \caption{Extrapolation of non-zero two-electron integrals with 10 basis functions per DG element.}
    \label{fig:nnz_tei_corr_calc_extrapolation}
\end{figure}

Figures~\ref{fig:H2_Gausslet_PES},~\ref{fig:H4_Gausslet_PES},~\ref{fig:H6_Gausslet_PES} show the DG calculations for H$_2$, H$_4$ and H$_6$ using Gausslets as the primitive basis set.
Table~\ref{tab:Gausslet_basis_set} shows information on the real space discretization for the used Gausslet basis set.
\begin{table}[h!]
\centering
\begin{tabular}{c|cccc}
\toprule
Atoms & H$_2$ & H$_4$ & H$_6$ & H$_8$ \\
\hline
tot. no. of Gausslet & 529 & 801 & 1{,}066 & 1{,}336\\
\hline
Gausslets in z-axis & 11-13 & 15-21 & 17-27 & 21-33\\
\hline
Gausslets in x- \\
resp. y-axis & 6.33-7.40 & 6.92-7.53 & 7.31-7.78 & 7.27-8.01\\
\hline
\end{tabular}
\caption{Averaged numbers Gausslets used to discretize the molecular system.}
\label{tab:Gausslet_basis_set}
\end{table}
Table~\ref{tab:numerical_values_Gausslets} compares the non-zero two-electron integral count for the DG-Gausslet basis and the active space basis. 
\begin{table}[h!]
\centering
\begin{tabular}{c|cc|cc}
\toprule
Atoms & H$_2$ & H$_4$ & H$_6$ & H$_8$ \\
\hline
No of Gausslets & 529 & 801 & 1{,}066 & 1{,}336 \\
\hline
nnz-tei Gausslet & 351{,}084 & 816{,}232 & 1{,}459{,}901 & 2{,}310{,}318\\
\hline
DG  per elem. & 6 & 6 & 10 & 10\\
nnz-tei DG & 4{,}144 & 16{,}580  & 287{,}762  &  511{,}449 	 \\
\hline
CAS & 4 & 8 & 12 & 16\\
nnz-tei CAS & 205  & 3{,}269 & 16{,}551  & 52{,}346 \\
\hline
\hline
\end{tabular}
\caption{Averaged numerical values of non-zero two-electron integrals for H$_2$, H$_4$, H$_6$ and H$_8$.}
\label{tab:numerical_values_Gausslets}
\end{table}
Figure~\ref{fig:nnz_tei_corr_calc_extrapolation} shows a naive extrapolation of the non-zero two-electron integral count presented in Table~\ref{tab:numerical_values_Gausslets}.

\section{Swap network sub-circuits}\label{sec:swap-details}

\begin{figure*}[t!]
    \centering
    \includegraphics[width=\textwidth]{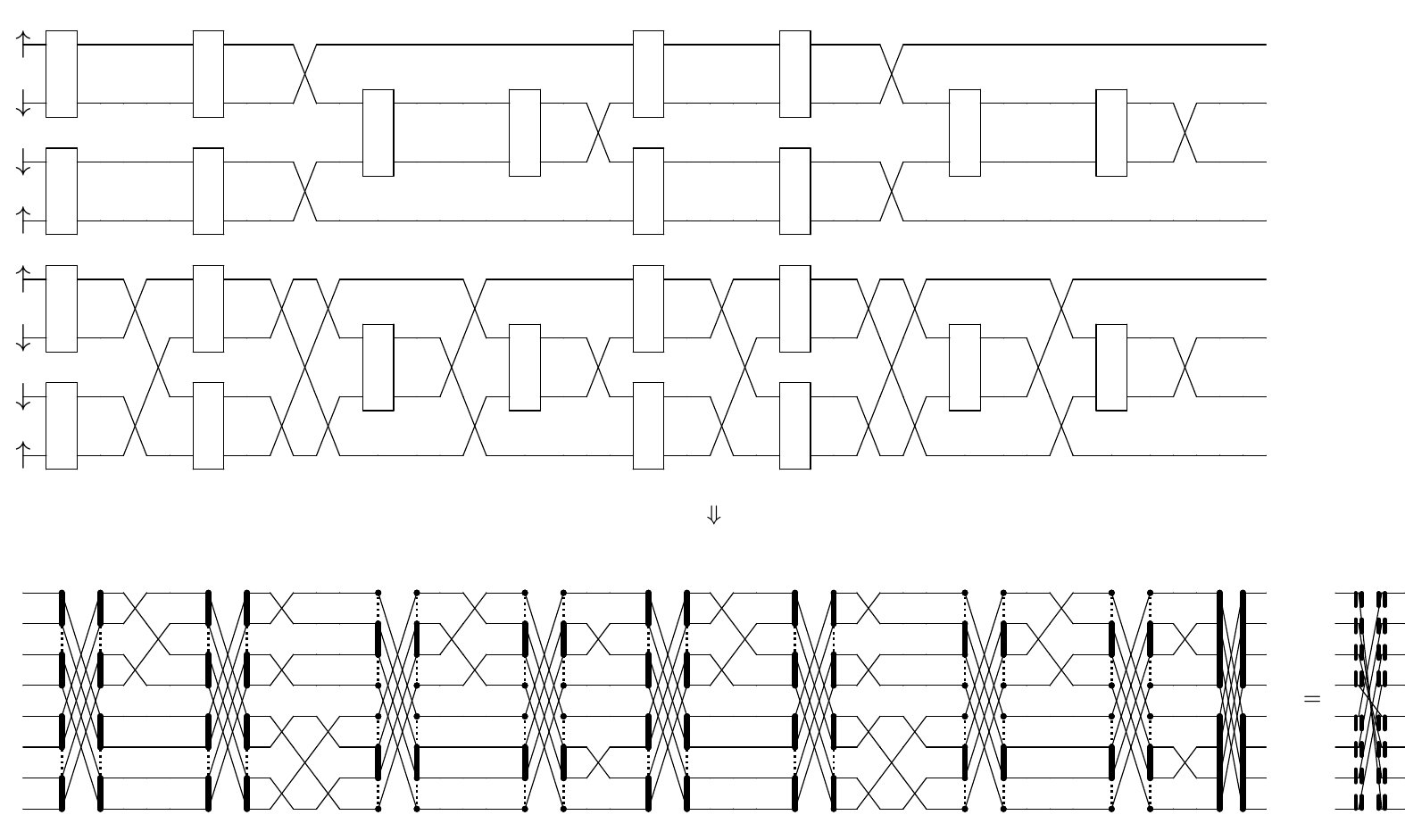}
    \caption{
Construction of the balanced double bipartite swap network.
Similar to the double bipartite swap network, except that pairs of orbitals from each part are only acquainted when their spins have the same parity.
The spins of the orbitals in the initial mapping of qubits to orbitals are indicated in the top left.\label{fig:double-bipartite-balanced}}
\end{figure*}

In this section, we give some more detail about the components of the swap network described in Section~\ref{sec:swap}.
Recall the structure of the overall swap network:
\begin{enumerate}
\item 
    A 4-complete swap network within each half-block.
    This acquaints all sets of $4$ orbitals with the same spin and within each block.
\item A double bipartite swap network on each block.
    This acquaints all sets of $4$ orbitals with no net spin and within each block.
    Details of the construction are given in Figure~\ref{fig:double-bipartite}.
\item A permutation within each block. This changes the orbital to qubit mapping in preparation for the next stage.
\item Alternating layers of balanced double bipartite swap networks.
    Each balanced double bipartite swap network acquaints, for some pair of blocks, all sets of $4$ orbitals with an even number of each spin and with two orbitals from each block.
    The $N_b$ alternating layers ensure that every pair of blocks is involved together in some balanced double bipartite swap network.
    Details of the construction of a double bipartite swap network are given in Figure~\ref{fig:double-bipartite-balanced}.
\end{enumerate}

\onecolumngrid

\begin{figure*}[t!]
    \centering
    \includegraphics{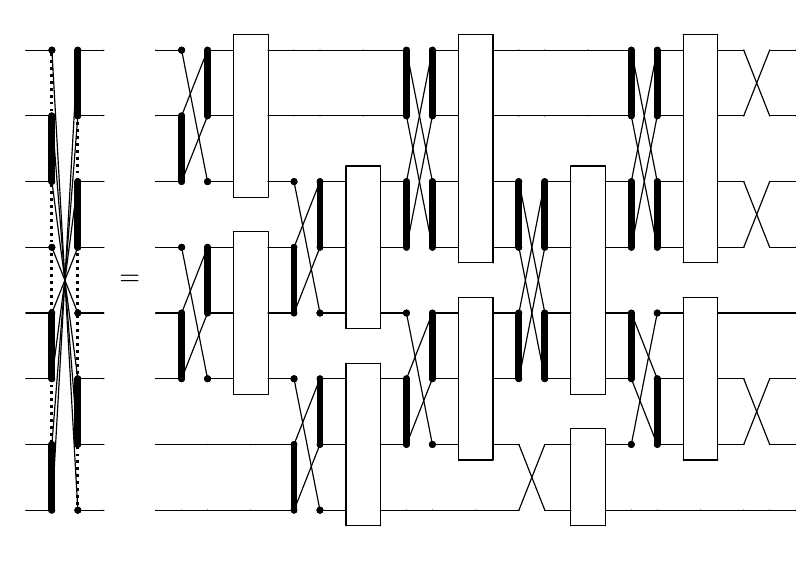}
    \caption{
Notation and decomposition for a $\mathcal P$-swap network with partition sizes $(1,2,1,2,2)$.
A $\mathcal P$-swap network for a partition $(P_1, P_2, \ldots, P_{|\mathcal P|})$ of the qubits $\bigcup_{i} P_i$ acquaints every union of a pair of parts, i.e., $\left\{P \cup P' | P, P' \in \mathcal P\right\}$.
At a high level, the structure is similar to that of the simple linear swap network, except that instead of single qubits being swapped, groups are (i.e., the parts of the partition $\mathcal P$).
There are $|\mathcal P|$ layers of generalized swap gates, each of which swaps sets of qubits.
For more details, see~\cite{ogorman2019generalized}.
\label{fig:partition-swap-network}}
\end{figure*}

\begin{figure*}[t!]
    \centering
    \includegraphics[width=\textwidth]{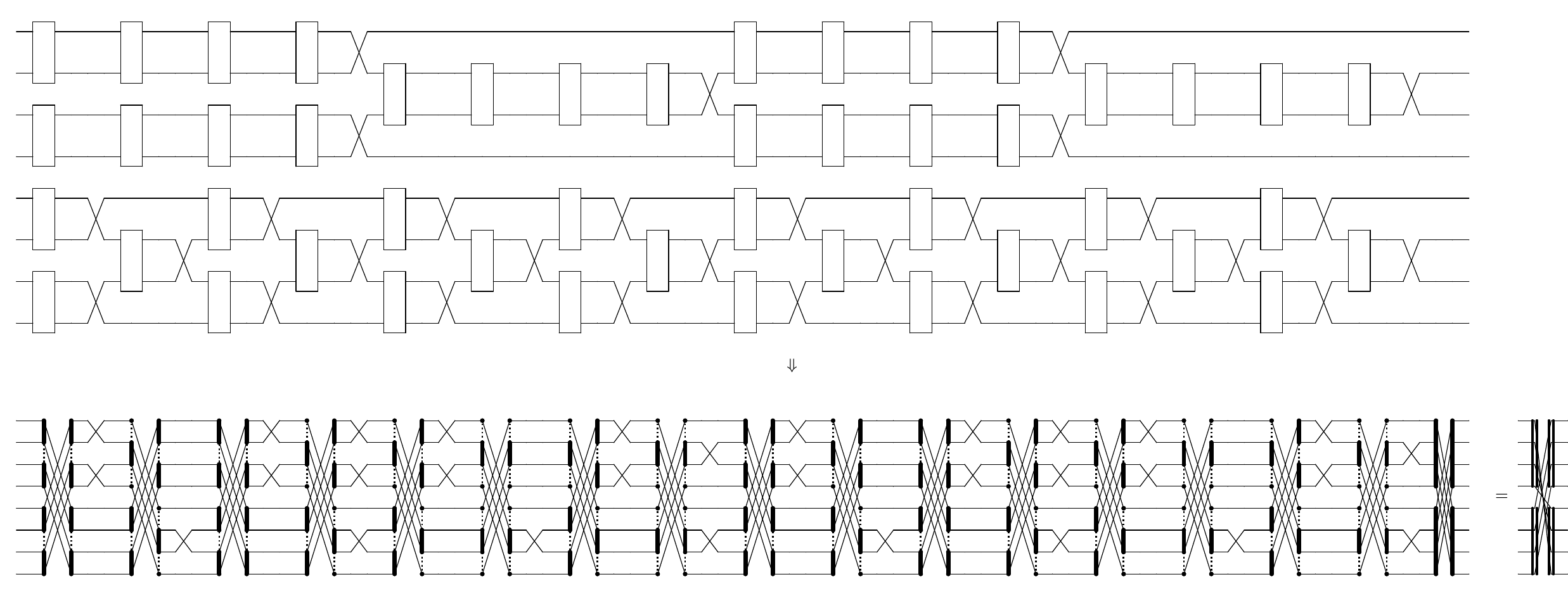}
    \caption{Construction of the double bipartite swap network, with parts of size $4$.
The top half of the top circuit contains the same swap gates as a linear swap network but with additional acquaintance opportunities.
In the bottom half of the top circuit are $4$ linear swap networks in a row, one for each acquaintance layer of the linear swap network in the top half, which is copied for each acquaintance layer of the bottom half.
Overall, for every set of of four orbitals consisting of two from the top part and two from the bottom part, there is a layer in the circuit in which both pairs are simultaneously acquainted.
The bottom circuit, depicting the double bipartite swap network, is formed by replacing each such acquaintance layer in the top circuit with a $\mathcal P$-swap network, where a pair of qubits acquainted in the top circuit corresponds to a part of the partition $\mathcal P$.
The $\mathcal P$-swap network acquaints the union of each pair of pairs; see Fig.~\ref{fig:partition-swap-network}.
The final gate ensures that overall effect is to shift the parts.
\label{fig:double-bipartite}}
\end{figure*}

\twocolumngrid

\end{document}